\documentclass[apj]{emulateapj-rtx4}
\usepackage{apjfonts}
\usepackage{times}
\usepackage{lscape}
\usepackage{amsmath}
\usepackage{graphicx}
\usepackage{natbib}
\usepackage{CJK} 
\usepackage{hyperref}

\shorttitle{Calibrating \ion{C}{4}-based $M_{\rm BH}$ estimators}
\shortauthors{Park et al.}

\newcommand{\mbh}{M$_{\rm BH}$}
\newcommand{\msun}{$M_{\odot}$}
\newcommand{\msigma}{$M_{\rm BH}-\sigma_{*}$}
\newcommand{\kms}{km~s$^{\rm -1}$}
\newcommand{\ergs}{erg~s$^{\rm -1}$}
\newcommand{\Hb}{H$\beta$}

\newcommand{\CIV}{\ion{C}{4}}
\newcommand{\MgII}{\ion{Mg}{2}}
\newcommand{\HST}{{\it HST}}
\newcommand{\IUE}{{\it IUE}}

\begin{document}
\begin{CJK*}{UTF8}{mj}

\title{Calibrating \ion{C}{4}-based Black Hole Mass Estimators}
\author{Daeseong Park (박 대 성 )$^{1}$}
\author{Jong-Hak Woo (우 종 학 )$^{1,\dagger}$}
\author{Kelly D. Denney$^{2}$}
\author{Jaejin Shin (신 재 진 )$^{1}$}
\affil{$^{1}$Astronomy Program, Department of Physics and Astronomy, Seoul National University, Seoul, 151-742, Republic of Korea; pds2001@astro.snu.ac.kr, woo@astro.snu.ac.kr, jjshin@astro.snu.ac.kr}
\affil{$^{2}$Dark Cosmology Centre, Niels Bohr Institute, Juliane Maries Vej 30, 2100 Copenhagen \O, Denmark; kelly@dark-cosmology.dk}
\altaffiltext{$\dagger$}{Author to whom correspondence should be addressed.}
%
%
\begin{abstract}
We present the single-epoch black hole mass estimators based on the \CIV~$\lambda$1549 broad 
emission line, using the updated sample of the reverberation-mapped AGNs and high-quality 
UV spectra. By performing multi-component spectral fitting analysis,
we measure the \CIV\ line widths (FWHM$_{\rm CIV}$ and line dispersion, $\sigma_{\rm CIV}$) 
and the continuum luminosity at 1350 \AA~($L_{1350}$) to calibrate the \CIV-based mass estimators.
By comparing with the \Hb\ reverberation-based masses, we provide new mass estimators 
with the best-fit relationships, i.e., $M_{\rm BH} \propto L_{1350}^{0.50\pm0.07} \sigma_{\rm CIV}^{2}$ and 
$M_{\rm BH} \propto L_{1350}^{0.52\pm0.09}{\rm FWHM}_{\rm CIV}^{0.56\pm0.48}$.
The new \CIV-based mass estimators show significant mass-dependent systematic difference 
compared to the estimators commonly used in the literature.
Using the published Sloan Digital Sky Survey QSO catalog, we show that 
the black hole mass of high-redshift QSOs decreases on average by $\sim 0.25$ dex if our recipe is adopted.
\end{abstract}

\section{INTRODUCTION}

Understanding the growth history of supermassive black holes (SMBHs) is one of the fundamental issues in
studies of galaxy formation and evolution.
The intimate connection between SMBHs and host galaxies is evidenced through
empirical correlations between the masses of SMBHs (\mbh) and the overall properties of the host galaxy spheroids
(e.g., Magorrian et al. 1998; Ferraresse \& Merritt 2000; Gebhardt et al. 2000).
The cosmic evolution of these scaling relationships has been investigated in the literature, where a 
tentative evolution has been reported utilizing observational approaches
(e.g., Peng et al. 2006; Woo et al. 2006, 2008; Treu et al. 2007; Merloni et al. 2010;
Bennert et al. 2010, 2011; Hiner et al. 2012; Canalizo et al. 2012).
In order to provide better empirical constraints on the cosmic growth of SMBHs and its connection to galaxy evolution,
reliable \mbh\ estimation at low and high redshifts is of paramount importance.

The \mbh\ can be determined for Type 1 AGN with the reverberation mapping (RM, Peterson 1993) method or 
the single-epoch (SE, Wandel et al. 1999) method under the virial assumption: $M_{\rm BH} = f R_{\rm BLR} {\varDelta V}^2/G$, where $G$ is the gravitational constant.
The size of the broad-line region (BLR), $R_{\rm BLR}$, can be directly measured from RM analysis
(e.g., Peterson et al 2004; Bentz et al. 2009; Denney et al. 2010; Barth et al. 2011b;
Grier et al. 2012)
or indirectly estimated from the monochromatic AGN luminosity measured from SE spectra
based on the empirical BLR size-luminosity relation (Kaspi et al. 2000, 2005; Bentz et al. 2006, 2009, 2013). 
The line-of-sight velocity dispersion, $\varDelta V$, of BLR gas can be measured either from the broad emission 
line width in the rms spectrum (e.g., Peterson et al. 2004) obtained from multi-epoch RM data or in the SE spectra (e.g., Park et al. 2012b), 
while the virial factor, $f$, is the dimensionless scale factor of order unity that depends on the geometry and kinematics 
of the BLR.
Currently, an ensemble average, $\langle f \rangle$, is determined empirically under 
the assumption that local active and inactive galaxies have the same \msigma~relationship (e.g., Onken et al. 2004; Woo et al. 2010; Graham et al. 2011; Park et al. 2012a; Woo et al. 2013) 
and recalibrated to correct for the systematic difference of line widths in between the SE and rms spectra (e.g., Collin et al. 2006; Park et al. 2012b).

The RM method has been applied to a limited sample ($\sim 50$) to date, due to the practical difficulty of the extensive photometric and spectroscopic monitoring observations and the intrinsic difficulty of tracing the weak variability signal across very long time-lags for high-z, high-luminosity QSOs.    
In contrast, the SE method can be applied to any AGN if a single spectrum is available, 
although this method is subject to 
various random and systematic uncertainties (see, e.g., Vestergaard \& Peterson 2006, Collin et al. 2006; McGill et al. 2008; Shen et al. 2008; Denney et al. 2009, 2012; 
Richards et al. 2011; Park et al. 2012b).

In the local universe, the SE mass estimators based on the \Hb\ line are well calibrated against the direct 
\Hb~RM results (e.g., McLure \& Jarvis 2002; Vestergaard 2002; Vestergaard \& Peterson 2006; Collin et al. 2006; Park et al. 2012b).
For AGNs at higher redshift ($z \gtrsim 0.7$), rest-frame UV lines, i.e., \MgII~or \CIV, are frequently used
for \mbh\ estimation since they are visible in the optical wavelength range.  
Unfortunately the kinds of accurate calibration applied to \Hb-based SE BH masses are difficult to achieve for 
the mass estimators based on the \MgII~and \CIV~lines, since the 
corresponding direct RM results are very few (see Peterson et al. 2005; Metzroth et al. 2006; Kaspi et al. 2007).
Instead, SE \mbh\ based on these lines can be calibrated indirectly against either the most reliable \Hb~RM based masses (e.g., Vestergaard \& Peterson 2006; Wang et al. 2009; Rafiee \& Hall 2011a) or the best calibrated \Hb~SE masses (McGill et al. 2008; Shen \& Liu 2012, SL12 hereafter)
under the assumption that the inferred \mbh\ is the same whichever line is used for the estimation.

While several studies demonstrated the consistency between \MgII~based  and \Hb\ based masses
(e.g., McLure \& Dunlop 2004; Salviander et al. 2007; McGill et al. 2008; Shen et al. 2008; Wang et al. 2009; Rafiee \& Hall 2011a; SL12), the reliability of utilizing the \CIV~line is still controversial, since \CIV\ can be severely 
affected by non-virial motions, i.e., outflows and winds, and strong absorption (e.g, Leighly \& Moore 2004; Shen et al. 2008; Richards et al. 2011; Denney 2012).
Other related concerns for the \CIV~line include the Baldwin effect, the strong blueshift or asymmetry of the line profile, broad absorption features, and the possible presence of a narrow line component (see Denney 2012 for discussions and interpretations of the issues).
Several studies have reported a poor correlation between \CIV~and \Hb~line widths and a large scatter between \CIV~and \Hb~based masses (e.g., Baskin \& Laor 2005; Netzer et al. 2007; Sulentic et al. 2007; SL12; Ho et al. 2012; Trakhtenbrot \& Netzer 2012).
On the other hand, other studies have shown a consistency between them and/or suggested additional calibrations 
for bringing \CIV~and \Hb~based masses further into agreement. (e.g., Vestergaard \& Peterson 2006; Kelly \& Bechtold 2007; Dietrich et al. 2009; Greene et al. 2010; Assef et al. 2011; Denney 2012).

Given the practical importance of the \CIV~line, 
which can be observed with optical spectrographs over a wide range of redshifts ($2 \lesssim z \lesssim 5$),
in studying high-z AGNs,
it is important and useful to calibrate the \CIV~based \mbh~estimators.
Vestergaard \& Peterson 2006 (VP06 hereafter) have previously calibrated \CIV\ mass estimators against
\Hb~RM masses, providing widely used \mbh\ recipes. Since then, however, the \Hb~RM sample has been expanded 
and many of RM masses have been updated based on various recent RM campaigns (e.g., Bentz et al. 2009; Denney et al. 2010; Barth 2011a,b; Grier et al. 2012).  At the same time, new UV data became available for the RM sample, 
substantially increasing the quality and quantity of available UV spectra for the RM sample.

In this paper we present the new calibration of the \CIV~based \mbh~estimators utilizing the highest quality UV
spectra and the most updated RM sample. In section 2 we describe the sample of \Hb~reverberation mapped AGN having available UV spectra.  Section 3 describes our detailed spectral analysis of the \CIV~emission line complex to obtain the relevant luminosity and line width measurements necessary for estimating SE \mbh. We provide the updated SE \CIV~\mbh~calibration in Section 4 and conclude with a discussion and summary in Section 5.
We adopt the following cosmological parameters to calculate distances in this work: 
$H_0 = 70$~km~s$^{-1}$~Mpc$^{-1}$, $\Omega_{\rm m} = 0.30$, and $\Omega_{\Lambda} = 0.70$.

%
%
\section{Sample and Data} \label{sec:sample}
%
%
%

For our analysis, we start with the reverberation mapped AGN sample, which is considered as a calibration base with 
reliable mass estimates. To date, there are $47$ AGNs, for which \Hb\ reverberation based masses are available 
(Peterson et al. 2004; Denney et al. 2006; Bentz et al. 2009; Denney et al. 2010; Barth et al. 2011a,b; Grier et al. 2012).
Of those $47$ objects, we selected $39$ AGNs\footnote{Note that we also excluded two objects in the list from Peterson et al. 2004 (i.e., PG 1211+143 and IC 4329A) due to the unreliable RM measurements as done in VP06, while we included NGC 4593 
since a new RM mass became available by Denney et al. (2006).} 
whose archival UV spectra are available from {\it International Ultraviolet Explorer} (\IUE) or {\it Hubble Space Telescope} (\HST) data archives\footnote{http://archive.stsci.edu/iue/ \& http://archive.stsci.edu/hst/}. 
First, we collected all available UV spectra covering the \CIV~spectral region from the public archives.
If there were multiple spectra for a given individual object, either multiple epochs taken with the same instrument or from multiple instruments,
we combined the spectra for each instrument by using a standard weighted average method to get the better signal-to-noise (S/N) ratio. At the same time, we tried to keep contemporaneity as far as possible.
Then we selected the best quality spectra for each object based on visual inspection and by setting a limiting S/N ratio of $\sim10$ per pixel, which was measured in an emission-line free region of the continuum near 1450\AA~or 1700\AA~ (see Denney et al. 2013 for the S/N related issues).
Among these $39$ objects, we excluded four AGNs (i.e., NGC 3227, NGC 4151, PG 1411+442, PG 1700+518) because they are severely contaminated with absorption features.
Other 9 AGNs (i.e., Mrk 79, Mrk 110, Mrk 142, Mrk 1501, NGC 4253, NGC 4748, NGC 6814, PG 0844+349, PG 1617+175) were also excluded due to the low quality and unreliability of UV spectra. 
Thus, our sample contains $26$ AGNs. Table~\ref{tab:optdata} lists the AGNs in the sample and their properties.
Note that we adopt the updated virial factor, $\log f = 0.71$ (Park et al. 2012a; Woo et al. 2013).

Compared to the previous sample of VP06, one object, Mrk 290, is newly included and 
seven objects (i.e., 3C 120, Mrk 335, NGC 3516, NGC 4051, NGC 4593, NGC 5548, PG 2130+099) have updated reverberation 
\mbh\ (Denney et al. 2006, Bentz et al. 2009; Denney et al. 2010; Grier et al. 2012). 
One object, PG 0804+761, that was excluded by VP06, is included since it has a new high-quality UV spectrum from the {\it Cosmic Origins Spectrograph} (COS) aboard \HST.
Contrary to VP06, NGC 4151 is omitted in this work due to the strong absorption features near the line center
(see Section 3).
In summary, $13$ AGNs have recent high-quality UV spectra from \HST~COS
\footnote{For the COS data, we performed a co-addition of multiple exposures with the upgraded costools routine (v2.0; Keeney et al. 2012) and then binned the spectra to a COS resolution element ($\sim0.07$ \AA, $\sim17$ \kms) by smoothing and re-binning by 7-pixels.} 
compared to VP06. For the remaining objects, UV spectra were obtained from the {\it Space Telescope Imaging Spectrograph} (STIS) aboard \HST\ for one object, from the {\it Faint Object Spectrograph} (FOS) aboard \HST\ for eight objects, and
from {\it Short-Wavelength Prime} (SWP) camera aboard \IUE\ for four objects as listed in Table~\ref{tab:UVdata}.
We corrected the Galactic extinction using the values of $E(B-V)$ from the recalibration of Schlafly \& Finkbeiner (2011) listed in
the NASA/IPAC Extragalactic Database (NED) and the reddening curve of Fitzpatrick (1999).

\section{Spectral Measurements} \label{sec:measurement}
%
%
%

\begin{figure*} 
\centering
\includegraphics[width=\textwidth]{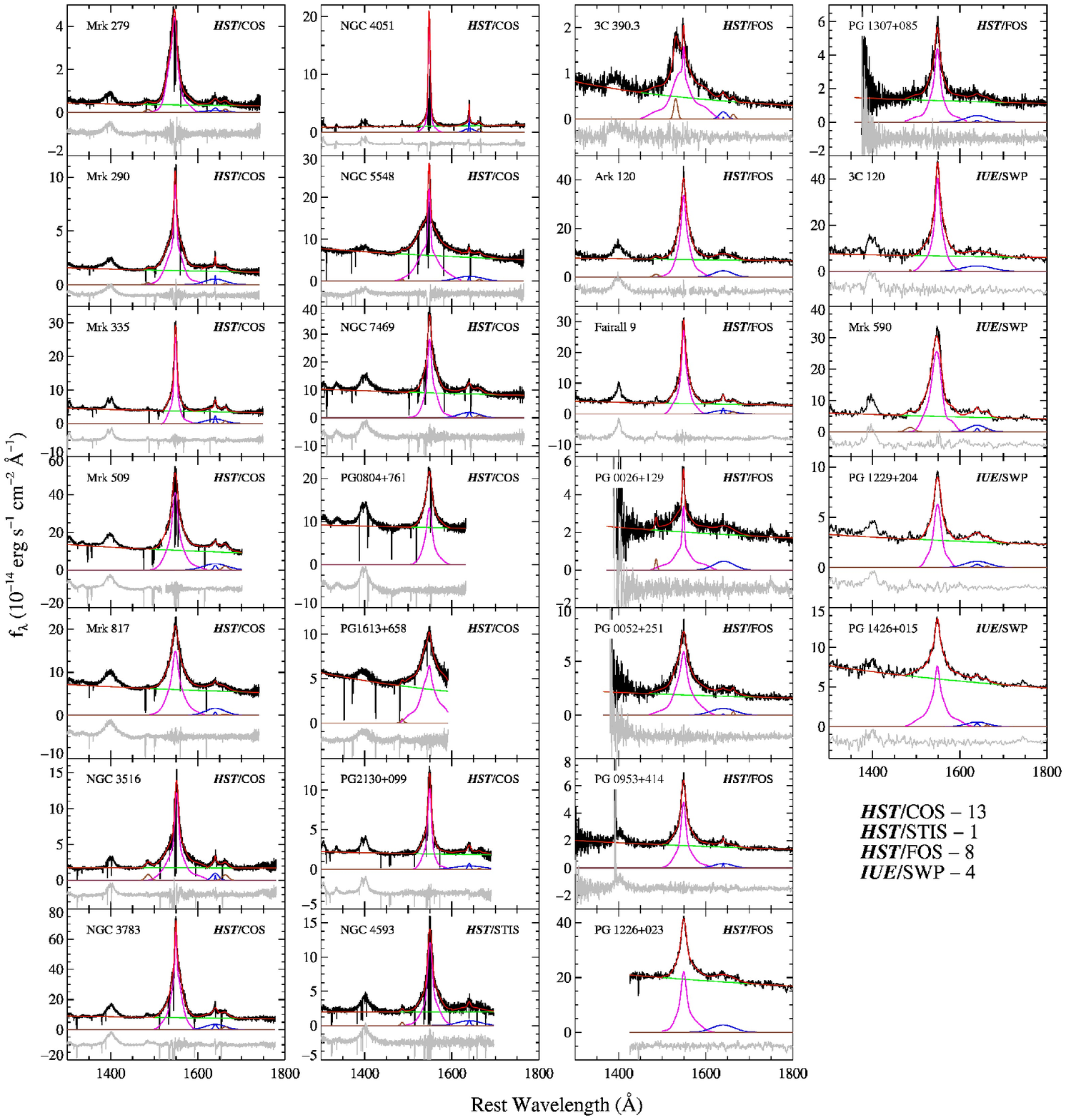}
\caption{
Multi-component fitting results for the sample of 26 objects.
In each panel, the observed UV spectrum (black) is overplotted with the best-fit model (red),
which consists of 
a single power-law continuum (green), 
\ion{C}{4} $\lambda$1549\AA~emission line (magenta), 
\ion{He}{2} $\lambda$1640\AA~emission line (blue), 
\ion{O}{3}] $\lambda$1663\AA~emission line (brown), and
\ion{N}{4}] $\lambda$1486\AA~emission line (brown).
The residual (gray), differences between black and red lines, is presented at the bottom of each panel.
}
\label{fig:all_spec_hst}
\end{figure*}

\begin{figure} 
\centering
\includegraphics[width=0.45\textwidth]{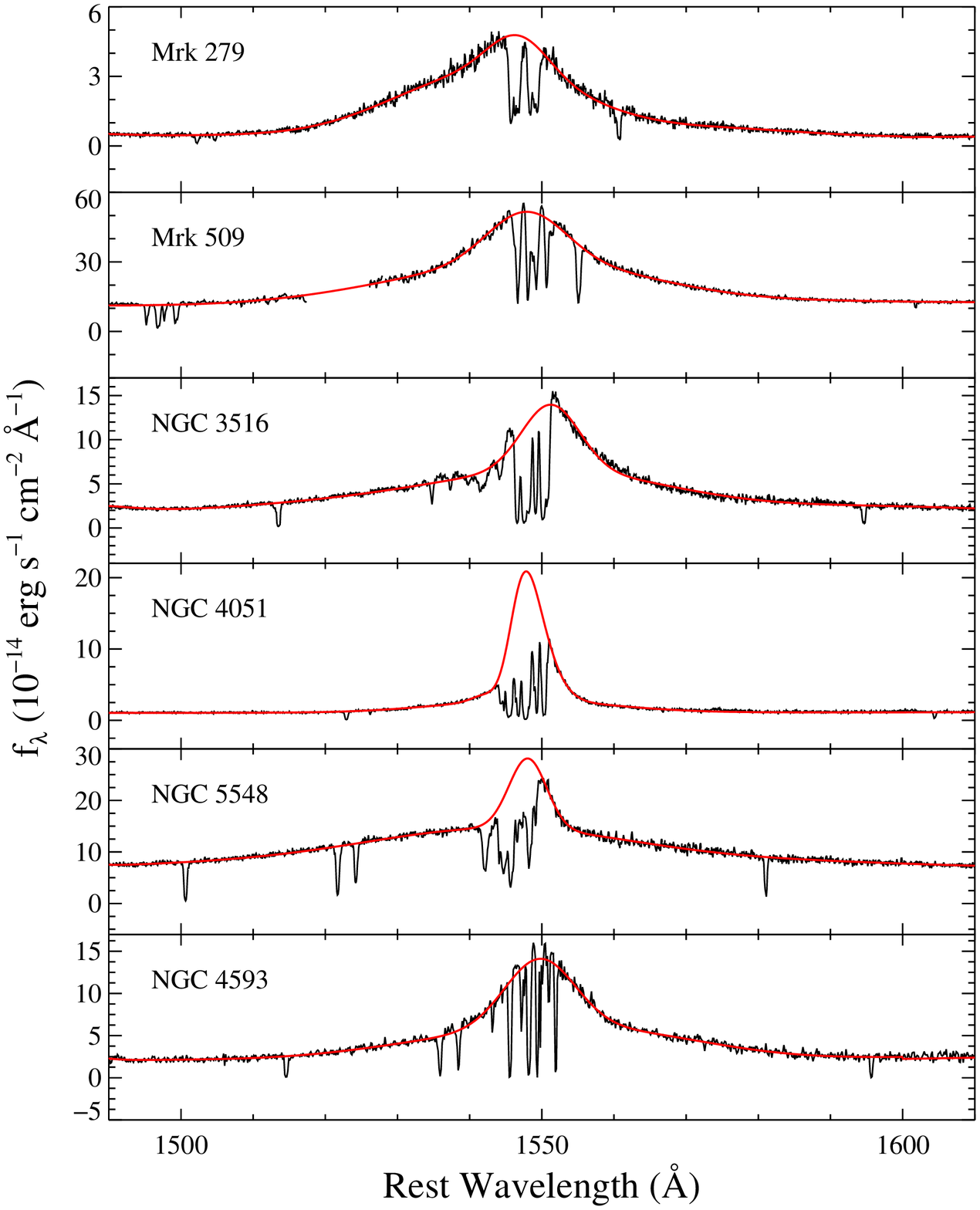}
\caption{
The zoom-in view of the \CIV\ for 6 objects, for which fitting results
are uncertain due to the absorption at the line center.
Black and red solid lines represent the observed spectra and the best-fit models, respectively.
}
\label{fig:obj_strabs}
\end{figure}
In order to calibrate the \CIV~\mbh\ estimator, we measured the line width of \CIV~ and
the continuum luminosity at 1350 \AA, following the multi-component fitting procedure developed by Park et al. (2012b) 
with a modification for the \CIV~region.
We first fitted a single power-law continuum using the typical emission-line-free windows in both sides of \CIV~
(i.e., $\sim 1340-1360$ \AA~or $\sim 1430-1470$ \AA~ and $\sim 1700-1730$ \AA), which were slightly adjusted for each 
spectrum to avoid the contaminating absorption and emission features.
We did not subtract the iron emission, since it is generally too weak to constrain at least in our data sets,
although we indeed tested the pseudocontinuum model by including the UV \ion{Fe}{2} template from Vestergaard \& Wilkes (2001).
After subtracting the best-fit power-law continuum, we simultaneously fitted the \CIV~complex region with the 
multi-component model consisting of a Gaussian function for the \ion{N}{4}] $\lambda$1486\AA, a Gaussian function for the \ion{Si}{2} $\lambda$1531\AA whenever clearly seen, a Gaussian function + a sixth-order Gauss-Hermite series for the \CIV~$\lambda$1549\AA, two Gaussian functions for the \ion{He}{2} $\lambda$1640\AA, and a Gaussian function for the \ion{O}{3}] $\lambda$1663\AA. 
Note that we fitted the 1600 \AA~feature, which is contaminating the red wing of \CIV, with a broad \ion{He}{2} 
component (cf. Appendix A. in Fine et al. 2010; Marziani et al. 2010).
In the fitting process,
the centers of \ion{He}{2}, \ion{O}{3}], \ion{N}{4}], and \ion{Si}{2} emission line components were fixed to be laboratory wavelengths.
We suppressed some components in \ion{He}{2}, \ion{O}{3}], \ion{N}{4}], and \ion{Si}{2} lines 
based on empirical tests with and without such components.
Narrow absorption features were excluded automatically in the calculation of $\chi^2$ statistics
by masking out the 3 sigma outliers below the smoothed spectrum (cf. Shen et al. 2011).
Strong broad absorption features around the \CIV~line center were also masked out manually 
by setting exclusion windows from visual inspection.

Although it is still controversial whether or not to remove a narrow emission-line component 
from \CIV before measuring the width, we use the full line profile of \CIV, i.e., without subtracting 
a narrow emission-line component, in order to be consistent with other studies (VP06, Shen et al. 2011, Assef et al. 2011; Ho et al. 2012).
We measured the continuum luminosities at $1350$ \AA~and $1450$ \AA~from the power-law model and measured 
the \CIV\ line widths (FWHM and $\sigma$) from the best-fit model (i.e., a Gaussian function + a sixth-order Gauss-Hermite series) as shown in Figure~\ref{fig:all_spec_hst}.
The measured line widths were corrected for the instrumental resolution following the standard practice 
by subtracting the instrumental resolution from the measured velocity in quadrature.
In Figure 2, we explicitly show spectra and best-fit models for the objects showing absorption features 
at the center of \CIV. Note that the fitting results are uncertain for these objects, 
in particular NGC 4051 (see Section 4.1).

To assess measurement uncertainties of the line width and continuum luminosity, 
we applied the Monte Carlo flux randomization method used by Park et al. (2012b; see also Shen et al. 2011).
Using the $1,000$ realizations of resampled spectra made by randomly scattering flux values based on the flux errors, 
we fitted and measured the line width and continuum flux, and
adopted the standard deviation of the distribution as the measurement uncertainties for individual objects
as listed in Table~\ref{tab:UVdata}.

\subsection{Continuum Luminosities and Line Widths}

\begin{figure} 
\centering
\includegraphics[width=0.45\textwidth]{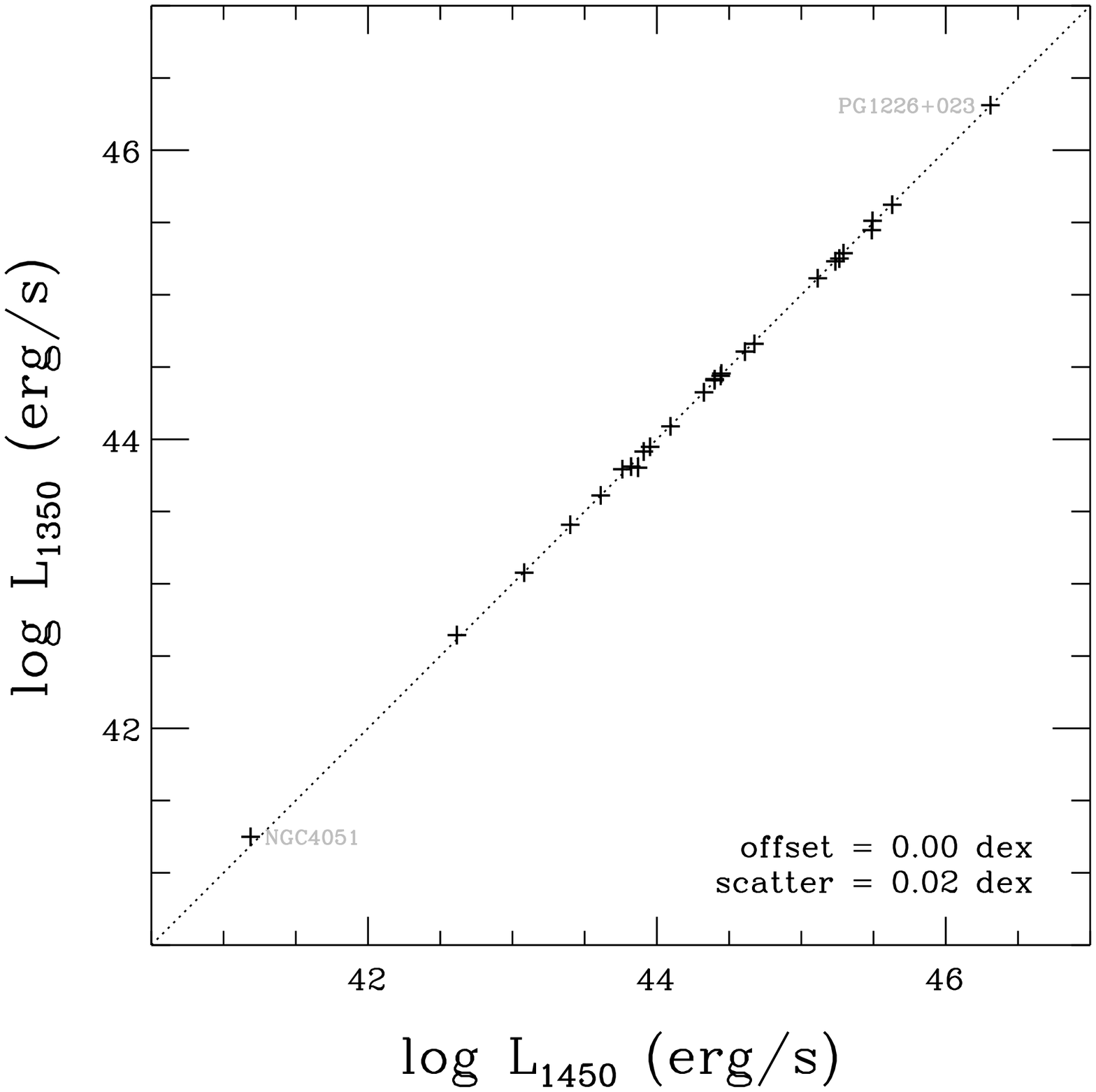} \\
\vspace{1em}
\includegraphics[width=0.45\textwidth]{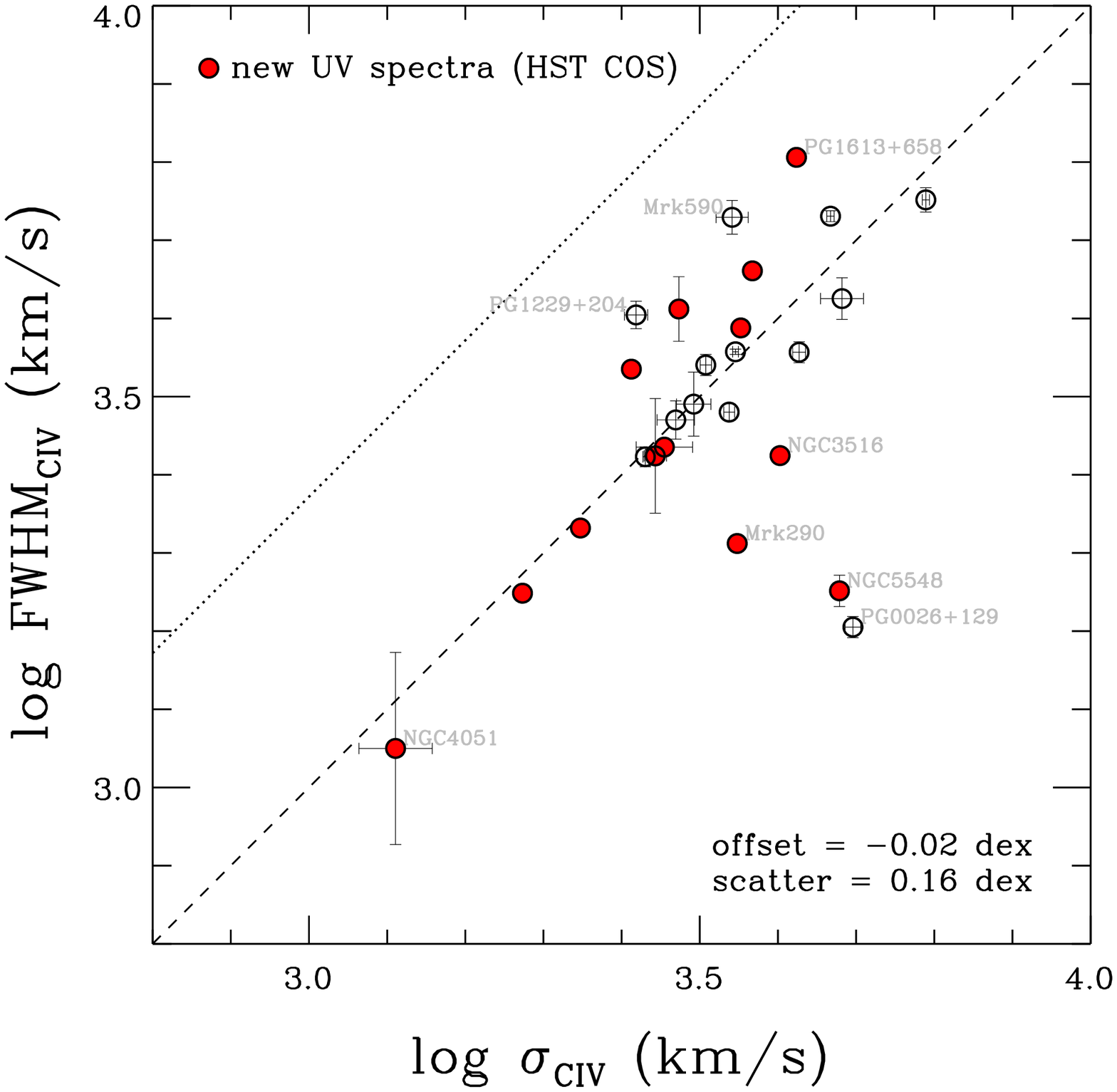}
\caption{
Comparison of UV measurements. The top panel shows the consistency of luminosities measured at $1350$ \AA~and $1450$ \AA,
respectively. 
The bottom panel compares the \CIV~FWHM and line dispersion ($\sigma$), where both were measured from the full line profile.
The ratio between FWHM and $\sigma$ is close to one (dashed line), indicating the line profile
is more peaky than Gaussian (dotted line).
Objects with new UV spectra from the \HST~COS is denoted with red filled circles.
Average offset and 1$\sigma$ scatter are given in the lower right corner in each panel.
}
\label{fig:UV_measurements}
\end{figure}

Figure~\ref{fig:UV_measurements} presents the continuum luminosities measured at $1350$ \AA~and $1450$ \AA, 
respectively, which are commonly adopted for the \CIV~\mbh~estimator.
Since they are almost identical, we choose to use $L_{1350}$ for the mass estimator.
The comparison between FWHM and $\sigma$ of \CIV\ is plotted in the bottom panel of Figure~\ref{fig:UV_measurements}.
It shows on average, a one-to-one relation between FWHM and $\sigma$, indicating that the \CIV\ profile is more peaky 
than a Gaussian profile, although there is large scatter.

\subsection{Comparison to Previous Measurements} \label{compare_VP06}
%
%
\begin{figure} 
\centering
\includegraphics[width=0.43\textwidth]{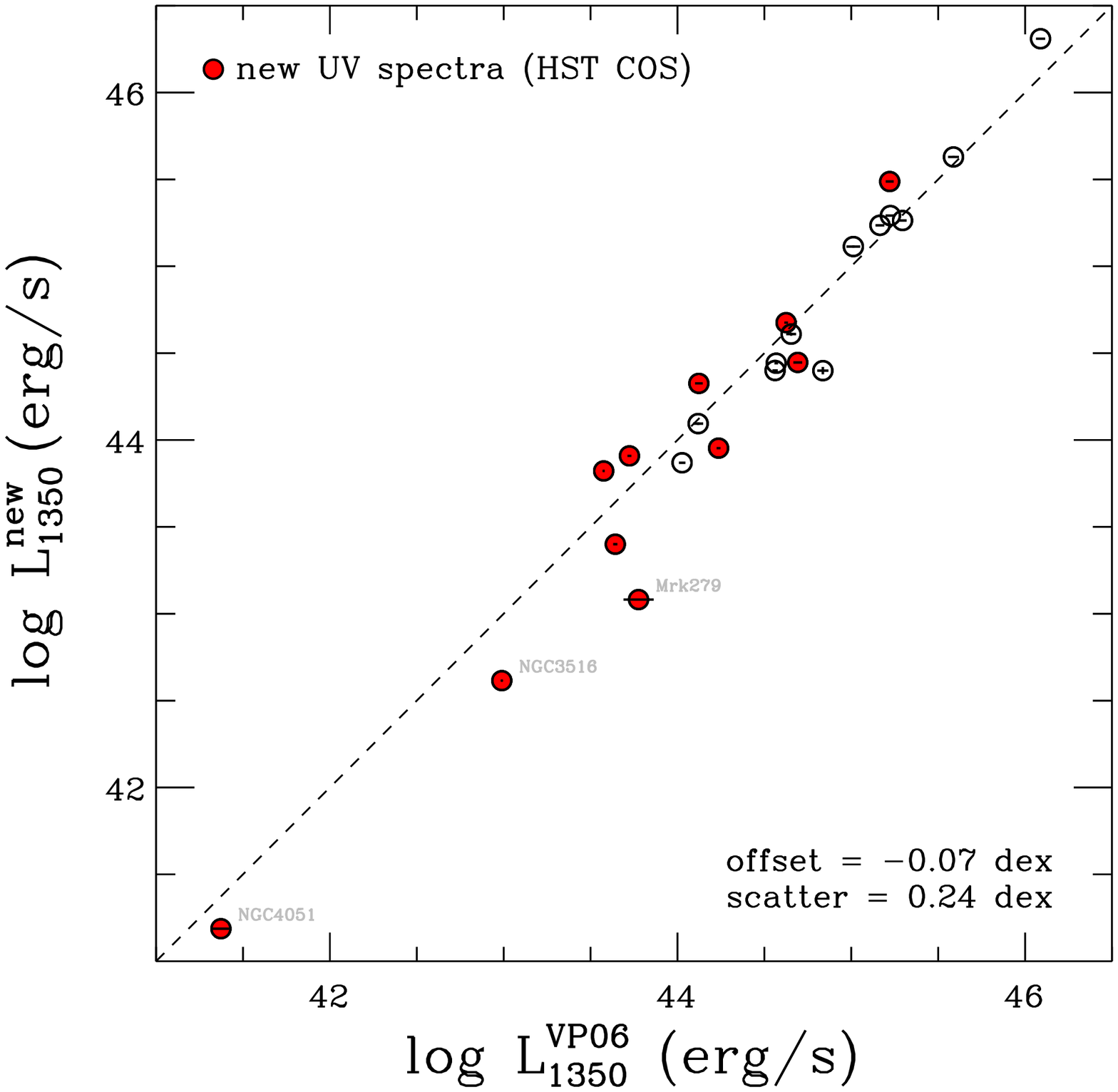} \\
\vspace{0.5em}
\includegraphics[width=0.43\textwidth]{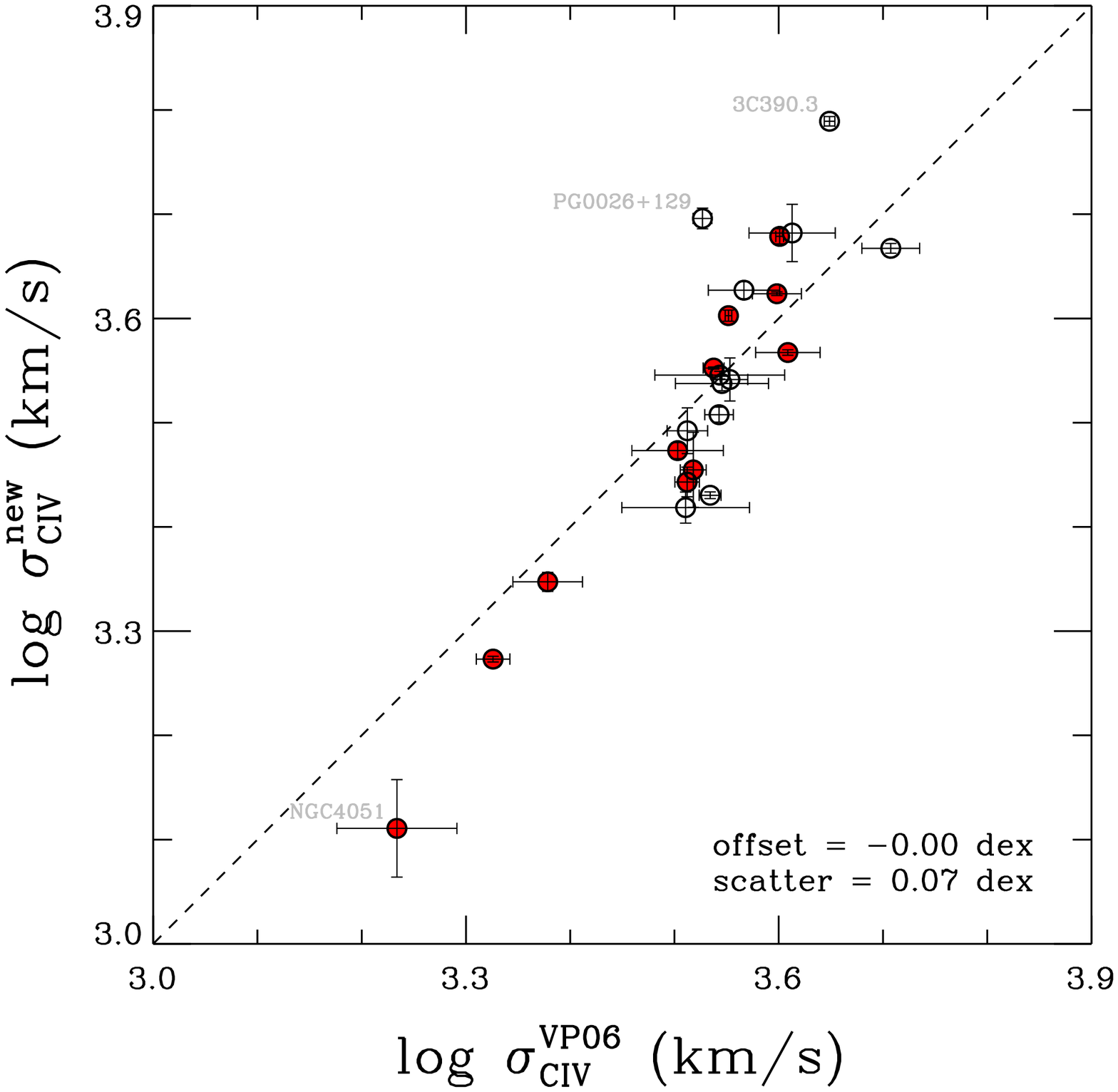} \\
\vspace{0.5em}
\includegraphics[width=0.43\textwidth]{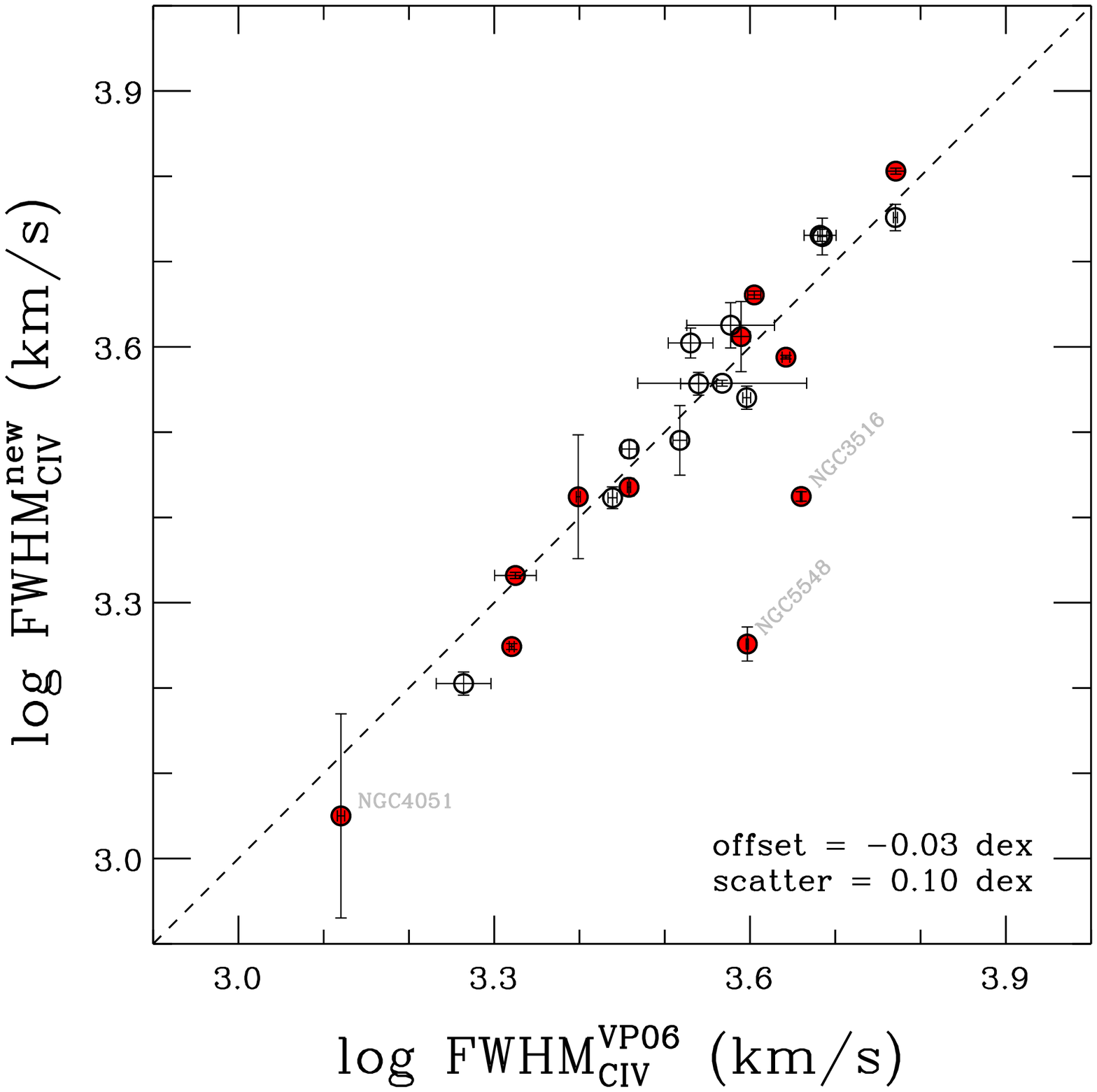}
\caption{
Comparisons of the luminosity (top) and line width (middle and bottom) measured in this study and those by VP06.
For VP06 values, the weighted average is plotted for given objects using the listed values in their Table 2.
Average offset and 1$\sigma$ scatter are given in the lower right corner in each panel.
}
\label{fig:comapre_UV_measure_VP06}
\end{figure}
We compare our measurements with those in VP06 in Figure~\ref{fig:comapre_UV_measure_VP06},
using the common sample (23 out of 27 objects given in their Table 2, except for Mrk 79, Mrk 110, NGC 4151, and PG 1617+175).
Since there are multiple measurements in VP06, we here show weighted average values of VP06 measurements for the purpose of comparison.

For the comparison of $L_{1350}$, there is 0.24 dex scatter, which may 
stem from a combination of the differences, e.g., adopted spectra and the Galactic extinction correction,
between our study and VP06.
We used the combined single spectra with the best quality while VP06 used all available SE spectra for each object.
Especially for the objects with the new \HST~COS spectra (red filled circles) observed in different epochs, 
there could be an intrinsic difference due to the variability.
In the case of the Galactic extinction correction, we utilized the recent values of $E(B-V)$ listed in the NED taken from the Schlafly \& Finkbeiner (2011) recalibration,
while VP06 used the original values from Schlegel et al. (1998).

When comparing our $\sigma_{\rm C IV}$ measurements to those of VP06, a slight positive systematic trend seems to be present (middle panel of Figure~\ref{fig:comapre_UV_measure_VP06}).
The most likely origin of this trend is the difference in the adopted line width measurement methods between VP06 and this work.
Based on the investigation by Fine et al. (2010), we modeled the \CIV~complex region with multiple components and measured line dispersion from 
the de-blended \CIV~line model profile, whereas VP06 measured line dispersion from the data without functional fits by limiting the \CIV~line profile range 
to $\pm10,000$ \kms~of the line center, regardless of the intrinsic line width of each CIV profile.
Thus, the line dispersion measured by VP06 will be biased if line wings are extended much further than the fixed line limit (i.e., underestimation) or
the wings are smaller than the fixed line limit (i.e., overestimation by including other features).
We avoid these biases by de-blending the \CIV~line from other lines using the multi-component fitting analysis and 
measuring the line widths from the best-fit models.
Since the line dispersion is more sensitive to the line wings than the line core, 
the decomposition and thus recovery of the line wing profile from contaminating lines is essential.

The bottom panel of Figure~\ref{fig:comapre_UV_measure_VP06} compares FWHM$_{\rm C\ IV}$, indicating 
on average consistency between VP06 and this work, except for a few outliers.
This is because FWHM is less sensitive to the line wings than $\sigma_{\rm C\ IV}$, hence the difference in the measuring
method does not generate significant difference in measurements. 
Note that although FWHM is sensitive to the narrow-line component, both VP06 and our study used the full line profile 
without decomposing the broad and narrow components.
Instead, another source of discrepancy comes from the fact that VP06 measured the line width directly from the data
while in this study the best-fit models were used for line width measurements. 
Thus, there may be object-specific differences depending on how the absorptions above the "half-maximum" flux level 
were dealt with by VP06, and how well the functional fits represent the peak of the profile in our study.

\section{Updating the Calibration of the \ion{C}{4} SE $M_{\rm BH}$ estimator} \label{sec:calibration}

\begin{figure} 
\centering
\includegraphics[width=0.45\textwidth]{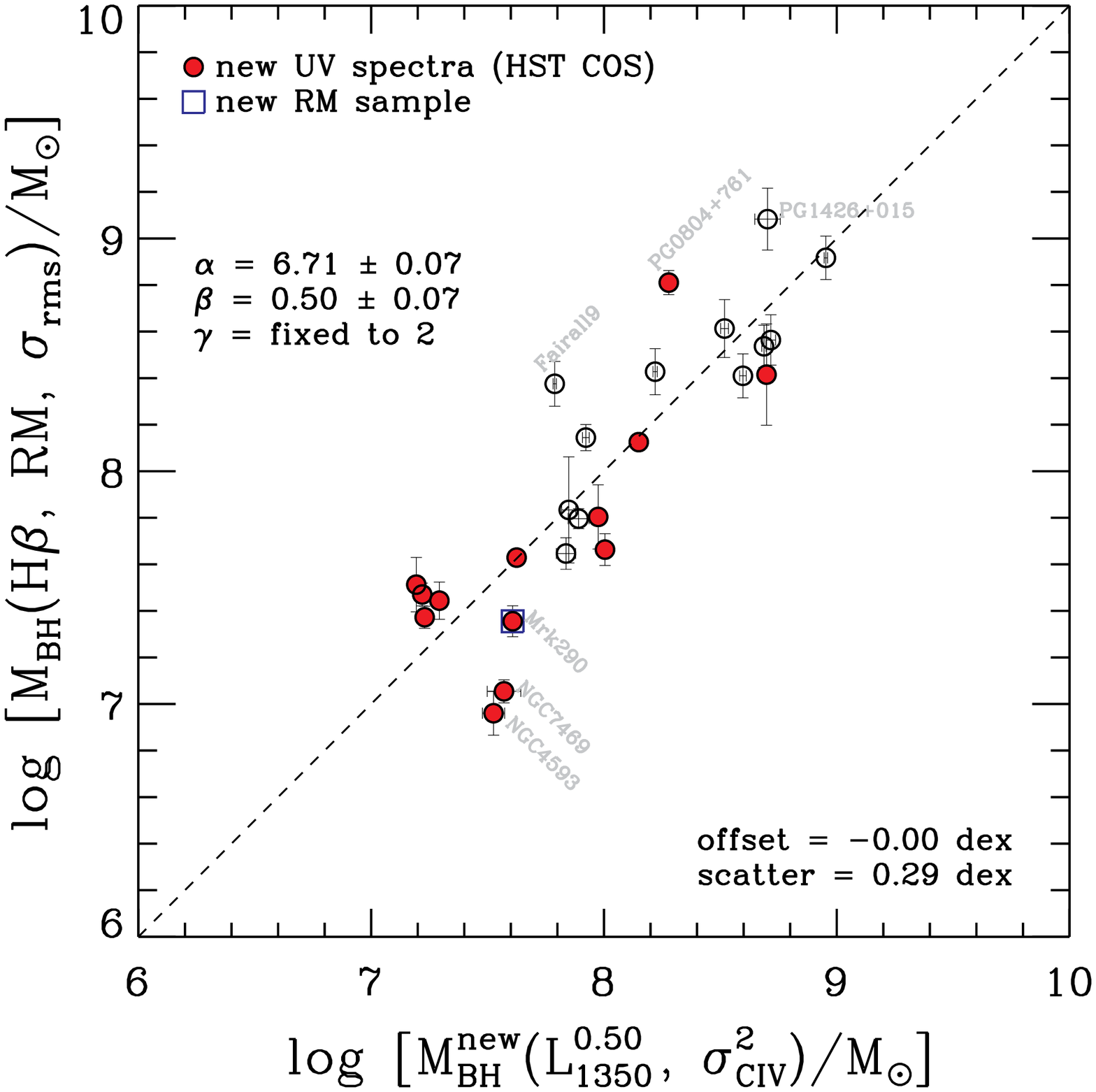} \\
\vspace{1em}
\includegraphics[width=0.45\textwidth]{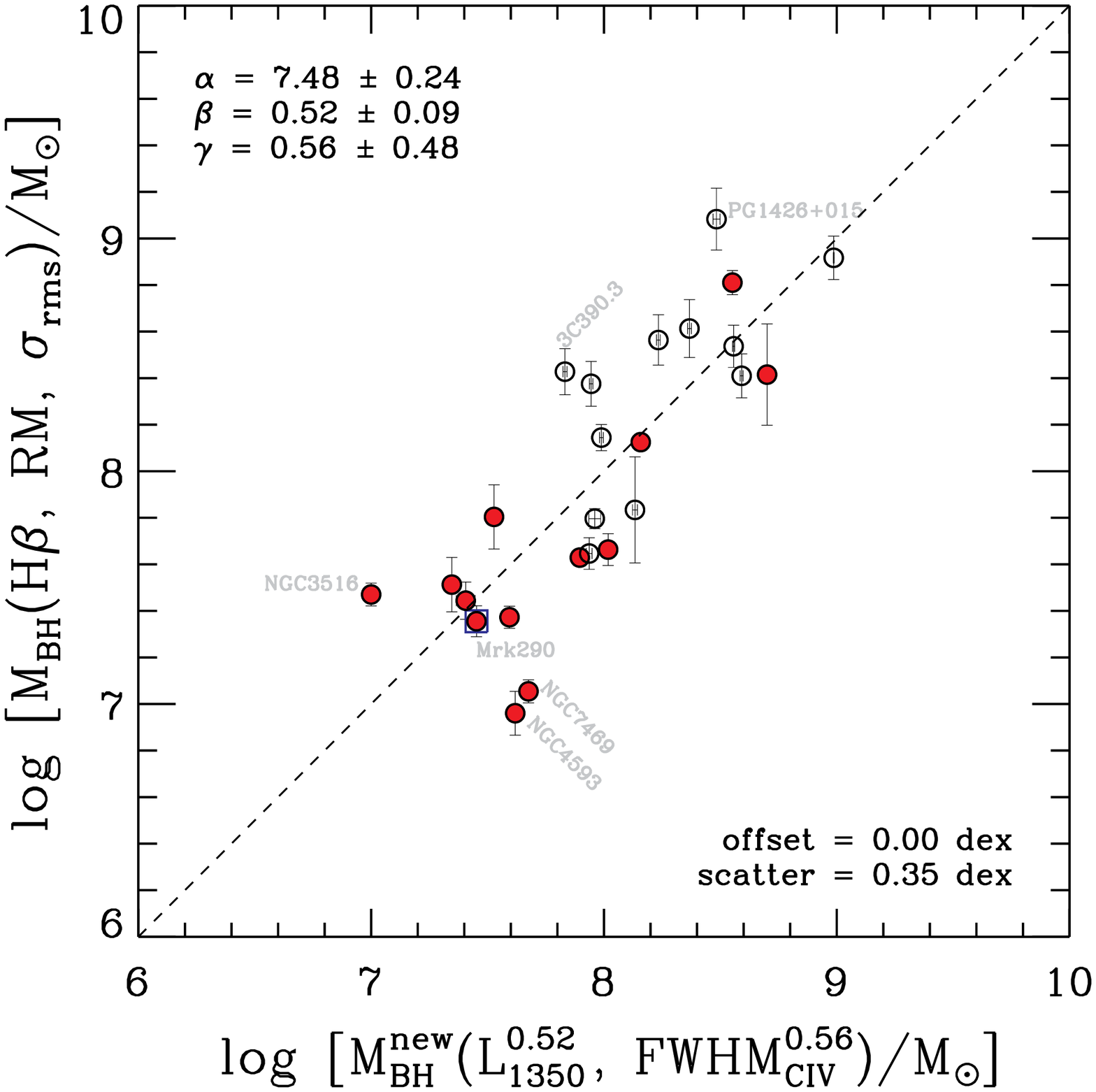} 
\caption{
Calibrations of $\sigma$-based SE BH masses (top) and FWHM-based SE BH masses (bottom) to the \Hb~RM-based BH masses.
The new sample from the recent RM results is marked with a blue open square.
The regressed parameters ($\alpha,\beta,\gamma$) with the uncertainty estimates are given in the upper left part in each panel.
}
\label{fig:recali3}
\end{figure}

By adopting \Hb~RM-based masses as true \mbh\ (see Table 1),
we calibrate the \CIV~mass estimators by fitting
\begin{eqnarray}\label{eq:calibration}
\log \left[\frac{M_{\rm BH} }{M_\odot}\right] ~=~  \alpha &+& \beta ~
\log\left(\frac{L_{1350}}{10^{44}~\rm erg~s^{-1}}\right) \nonumber\\
 &~+~&  \gamma ~ \log \left[\frac{\varDelta V(\textrm{\CIV})}{1000~ \rm km~s^{-1}}\right]~,
\end{eqnarray}
where $L_{1350}$ is the monochromatic continuum luminosity at 1350~\textrm{\AA} and 
$\varDelta V(\textrm{\CIV})$ is the line width of \CIV, either FWHM$_{\rm CIV}$ or 
$\sigma_{\rm CIV}$.
We regress Equation~\ref{eq:calibration} to determine the free parameters ($\alpha,\beta,\gamma$) using the \texttt{FITEXY} estimator implemented in Park et al. (2012a).
Note that this approach is different from that used by VP06, who adopted the luminosity slope from the size-luminosity relation and fixed the velocity slope to $2$.
Instead, this method is consistent with the recent approaches described by Wang et al. (2009), Rafiee \& Hall (2011b), and Shen \& Liu (2012).
Because a non-linear dependence is often observed between the line widths of \Hb~and \CIV~line (especially based on the FWHM; see Denney 2012 for a likely physical interpretation), leaving $\alpha$, $\beta$, and $\gamma$ as free parameters arguably results in a better statistical regression by accommodating the possible covariance between luminosity and line width.

\subsection{New Calibrations}

In Figure~\ref{fig:recali3}, we present the final best-fit calibration results for \CIV-based mass estimators
by directly comparing the \CIV~SE masses with the \Hb~RM-based masses, using Equation~\ref{eq:calibration}. 
The regression results with various conditions and the previous calibrations from the literature are listed in Table~\ref{tab:calibration}.
We adopt the regression results without NGC 4051, which is subject to the largest measurement errors among our sample
since  modeling the \CIV\ line of this object is highly uncertain due to the strong absorption at the center
(see Fig.~\ref{fig:obj_strabs}).
In addition, the variability of NGC 4051 is expected to have a large amplitude since it is the lowest-luminosity object
in our sample. Thus, NGC 4051 can add large scatter to the regression and potentially skew the slope because there is only a single object at the low-mass regime.
Thus, excluding this object possibly will lead to less biased results in terms of sample selection and measurement
uncertainties. We will present the results without NGC 4051 hereafter unless explicitly stated.

The slope of the velocity term, when it is treated as a free parameter, is closer to the virial assumption (i.e., 2) for the $\sigma_{\rm C\ IV}$-based estimator, i.e., $\gamma=1.74\pm0.55$ ($\gamma=1.45\pm0.53$ if NGC 4051 is included) than
for the FWHM$_{\rm C\ IV}$-based estimator, i.e., $\gamma=0.56\pm0.48$ ($\gamma=0.52\pm0.46$ if NGC 4051 is included).
This reinforces the use of $\sigma$ for characterizing the line width of \CIV, as suggested by Denney 
(2012, 2013; see also Peterson et al. 2004 and Park et al. 2012b for the case of \Hb).
The slope of the luminosity term (i.e., $\beta=0.51\pm0.08$ for $\sigma$; $\beta=0.52\pm0.09$ for FWHM) is almost 
consistent to that of photoionization expectation (i.e., 0.5; Bentz et al. 2006) within the uncertainty.
This may indicates that asynchronism between \Hb~and \CIV~measurements does not introduce a significant overall difference for our high-luminosity, high-quality calibration sample.

In this calibration, we treat $\beta$ and $\gamma$ as free parameters in addition to $\alpha$.
Letting $\beta$ be a free parameter is required to reduce luminosity dependent systematics since we are dealing with non-contemporaneous \Hb~and \CIV~measurements, which is expected to be not necessarily linear.
In addition, it is currently questionable to directly adopt the \CIV~size-luminosity relation (e.g., Kaspi et al. 2007) for the estimator 
since it is based on such a small sample.
Relaxing the constraint of $\gamma = 2$ for FWHM$_{\rm C\ IV}$ can be corroborated by the investigation by Denney (2012), which shows that there are severe biases in measuring FWHM$_{\rm C\ IV}$ due to the non-variable component and dependence on the line shape. These systematic uncertainties may be properly calibrated out by taking $\gamma$ as a free
parameter.
In the case of $\sigma_{\rm C\ IV}$, however,
a similar systematic does not seem to be present for $\sigma_{\rm C\ IV}$ (see Denney 2012). 
Even if we allow $\gamma$ to be free, the regression slope for $\sigma_{\rm C\ IV}$ is consistent to the virial
expectation (i.e., 2) within $1\sigma$ uncertainty, thus we opt to fixing $\gamma$ to a value of 2, 
avoiding systematic uncertainties due to small number statistics or sample-specific systematics.
Thus, here we provide the best estimator for the \CIV-based \mbh\ (see also, Fig.~\ref{fig:recali3}) as
\begin{eqnarray}\label{eq:final_cal_sigma}
\log \left[\frac{M_{\rm BH} }{M_\odot}\right] 
&~=~& (6.71\pm0.07) \nonumber\\
&~+~& (0.50\pm0.07) ~ \log\left(\frac{L_{1350}}{10^{44}~\rm erg~s^{-1}}\right) \nonumber\\
&~+~& 2 ~ \log \left[\frac{\sigma (\textrm{\CIV})}{1000~ \rm km~s^{-1}}\right]
\end{eqnarray}
with the statistical scatter against RM masses of $0.29$ dex and
\begin{eqnarray}\label{eq:final_cal_fwhm}
\log \left[\frac{M_{\rm BH} {\rm (SE)}}{M_\odot}\right] 
 &~=~& (7.48\pm0.24) \nonumber\\
 &~+~& (0.52\pm0.09) ~ \log\left(\frac{L_{1350}}{10^{44}~\rm erg~s^{-1}}\right) \nonumber\\
 &~+~& (0.56\pm0.48) ~ \log \left[\frac{\rm FWHM (\textrm{\CIV})}{1000~ \rm km~s^{-1}}\right]
\end{eqnarray}
with the statistical scatter against RM masses of $0.35$ dex.
Apart from the interpretation of values of zero point and slopes, it is worth noting that these estimators
are the best calibrated ones to reproduce \Hb~RM masses as closely as possible for the current sample and data 
sets.

\subsection{Comparison to Previous Recipes}

%
\begin{figure} 
\centering
\includegraphics[width=0.43\textwidth]{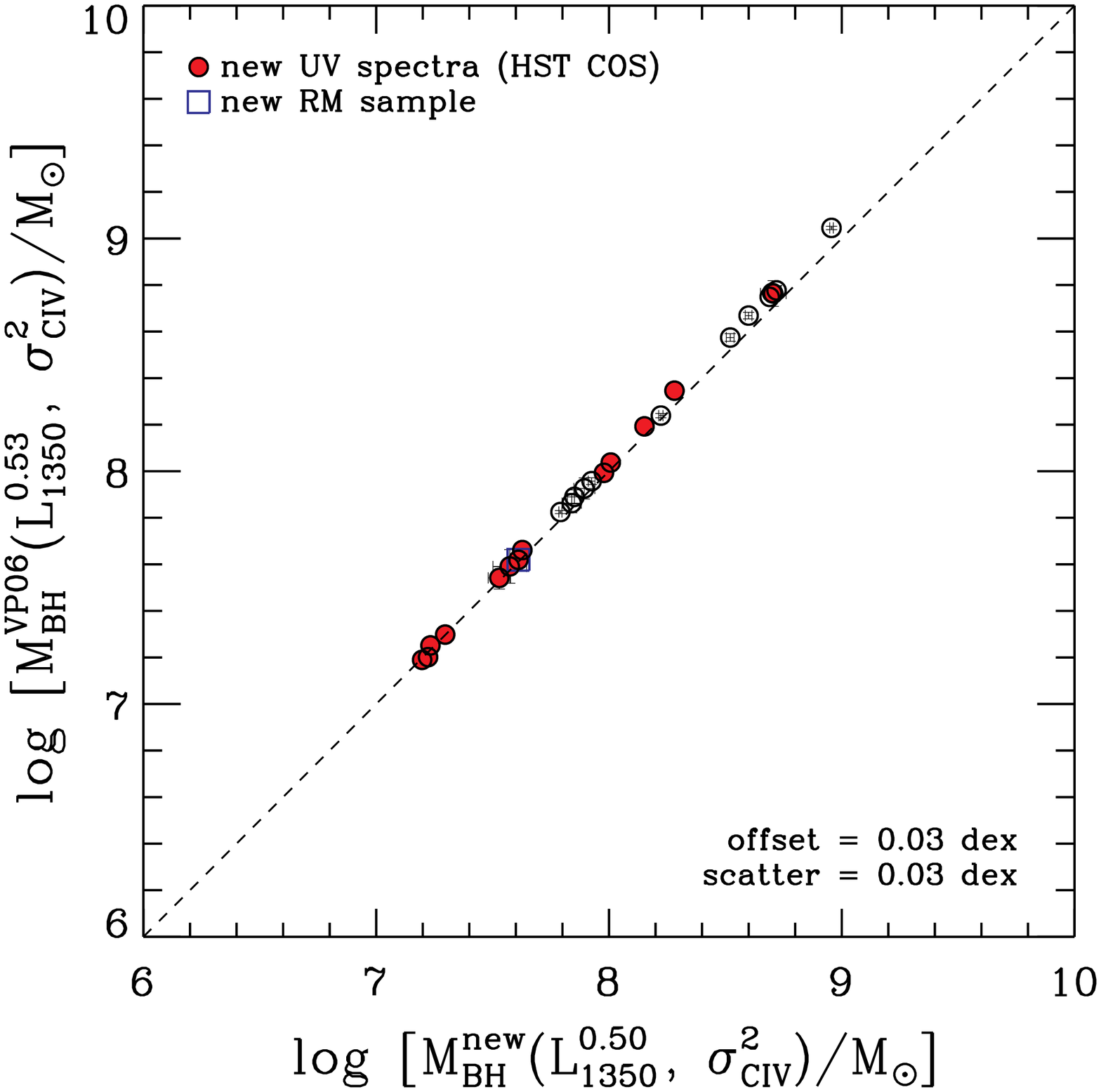} \\
\vspace{0.5em}
\includegraphics[width=0.43\textwidth]{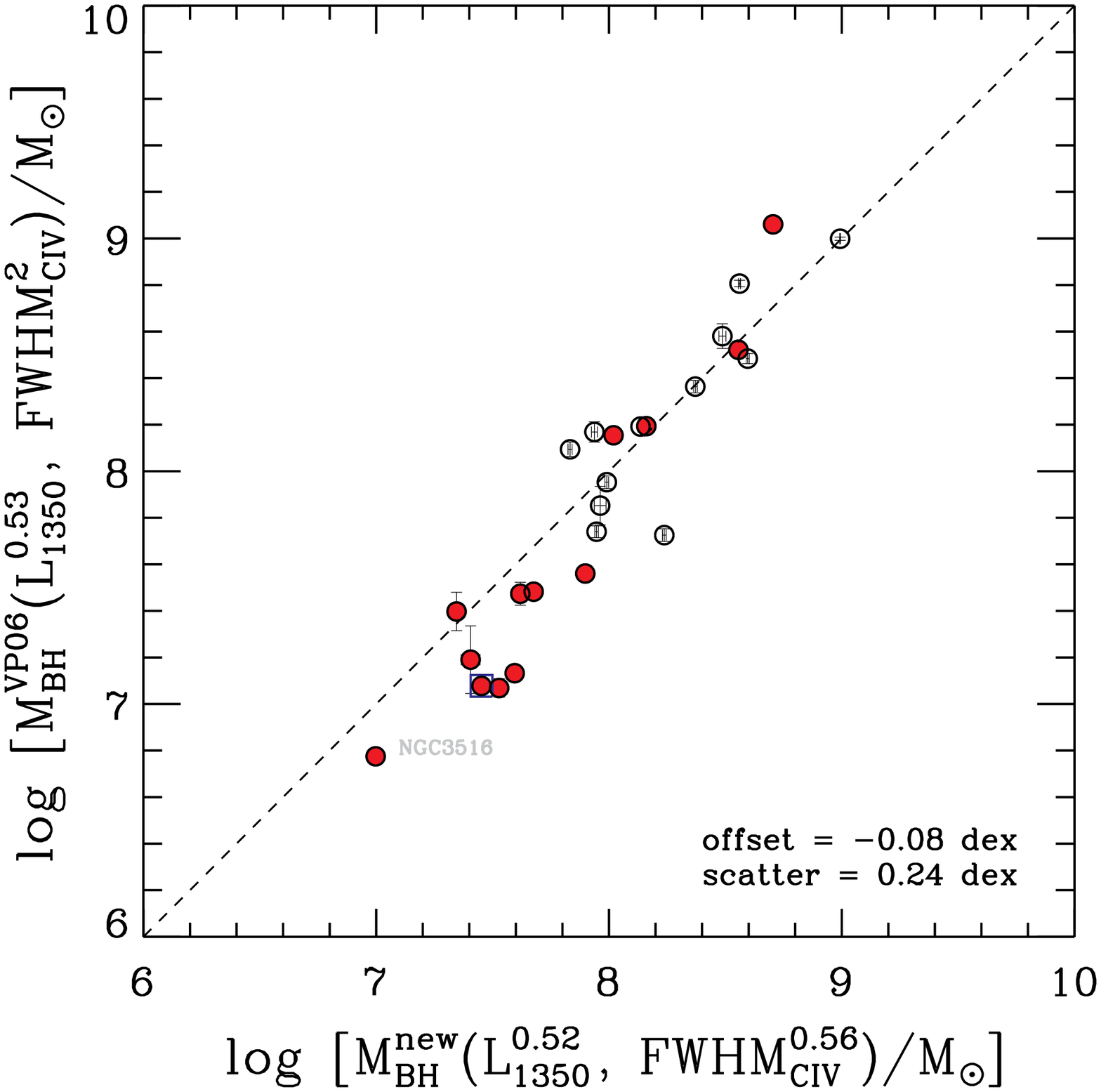} \\
\vspace{0.5em}
\includegraphics[width=0.43\textwidth]{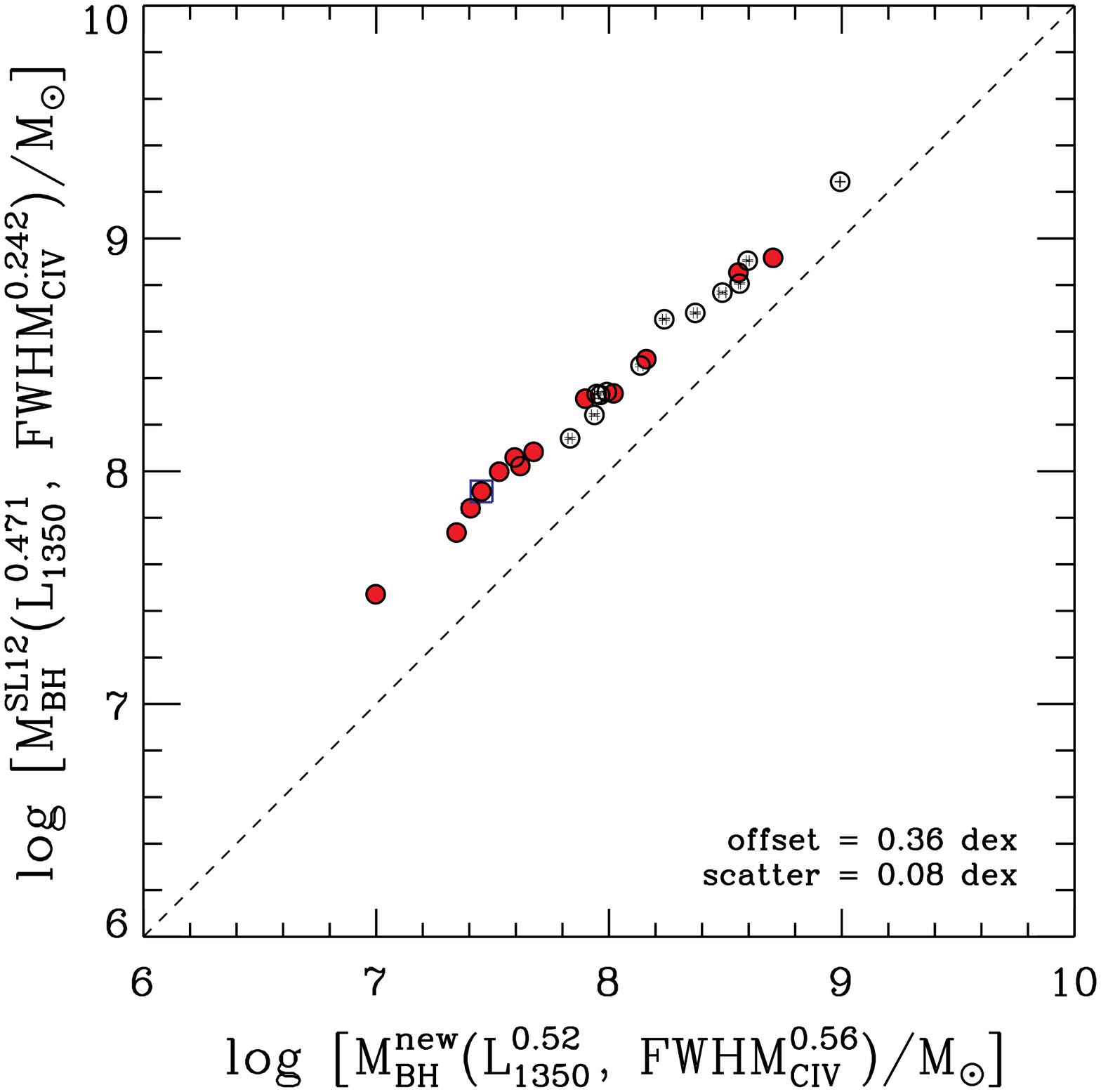}
\caption{
Comparison of $\sigma$-based (top) and FWHM-based (middle) \mbh\, respectively calculated using the estimators in this 
study and the estimators of VP06.
In the bottom panel, \mbh~calculated with the FWHM-based estimator from SL12 is compared to our mass estimates.
Average offset and 1$\sigma$ scatter are given in the lower right corner in each panel.
}
\label{fig:comapre_vp06_sl12}
\end{figure}

In Figure~\ref{fig:comapre_vp06_sl12}, we present the systematic difference of \mbh\ estimates based on our new
estimators (Equations~\ref{eq:final_cal_sigma} and \ref{eq:final_cal_fwhm}) compared to the previous estimators
from VP06 and SL12, respectively, using \CIV\ line width and L$_{1350}$ measurements. 
The $\sigma_{\rm C\ IV}$-based \mbh~estimates show almost consistent results to VP06 with a slight offset of $\sim0.03$,
which is expected from a difference in the adopted values of the virial factor (i.e., $\log f = 0.71$ here versus $\log f = 0.74$ in VP06).
In contrast, the comparison of FWHM$_{\rm C\ IV}$-based masses, respectively estimated with our recipe and with
that of VP06, shows large scatter and a systematic trend that the VP06 recipe underestimates \mbh~in the low-mass 
regime and overestimates \mbh~in the high-mass regime, compared to our recipe.
The bottom panel of Figure~\ref{fig:comapre_vp06_sl12} shows that the SL12 recipe systematically overestimates \mbh~over the whole dynamic range of the sample (i.e., $\lesssim 10^{9}$ \msun).
This is understandable because SL12 used the FWHM$_{\rm H\beta}$-based \mbh~in VP06 as a 
fiducial one and recalibrate the estimator using 
their high-mass ($>10^9$ \msun) QSOs. Thus, the calibration performed by SL12 in their limited dynamical range
inherits the overestimation behavior of the VP06 recipe with respect to our recipe, and 
propagates it into the low mass regime with larger effect.
In addition, SL12 subtracted a narrow \CIV~component before measuring FWHM of \CIV, leading to an overestimated FWHM$_{\rm C\ IV}$, compared to VP06 and our methods.  
We note that a large dynamic range is necessary for better calibration and investigation of the biases, 
as pointed out by SL12.

In order to explicitly compare the calibration methods used in here and VP06, we regress Equation~\ref{eq:calibration} 
by fixing $\beta$ and/or $\gamma$ with the adopted values in VP06 as listed in Table~\ref{tab:calibration}.
For the $\sigma$-based mass estimator, we obtain almost same calibration result ($\alpha=6.72\pm0.06$) to that of 
VP06 ($\alpha=6.73\pm0.01$) using the sample including NGC 4051.
When NGC 4051 is exclude, the zero points reduces slightly ($\alpha=6.69\pm0.06$) and intrinsic scatter becomes smaller.
It is interesting to see the consistency of the $\sigma$-based calibration between our study and VP06,
despite the systematic bias in $\sigma_{\rm C IV}$ measurements of VP06 as shown in Section~\ref{compare_VP06}.
We interpret this as follows. Although there is a bias in the VP06 measurement method for $\sigma$, 
due to their choice of line limits (i.e., $\pm10,000$ \kms), their $\sigma$-based \mbh~measurements serendipitously scatter evenly below and
above the central point of the mass scale of the sample, consequently resulting in a similar zero point 
regardless of the bias in $\sigma$ measurements. In the end the calibrations are very similar, however, 
the intrinsic scatter of our calibration is smaller than that of VP06, which demonstrates a general increase 
in accuracy of our $\sigma$-based masses over those of VP06, advocating for our $\sigma$ measurement prescription.

\subsection{Difference in SMBH Population using the SDSS DR7 Quasar Catalog}

\begin{figure*} 
\centering 
\includegraphics[width=0.8\textwidth]{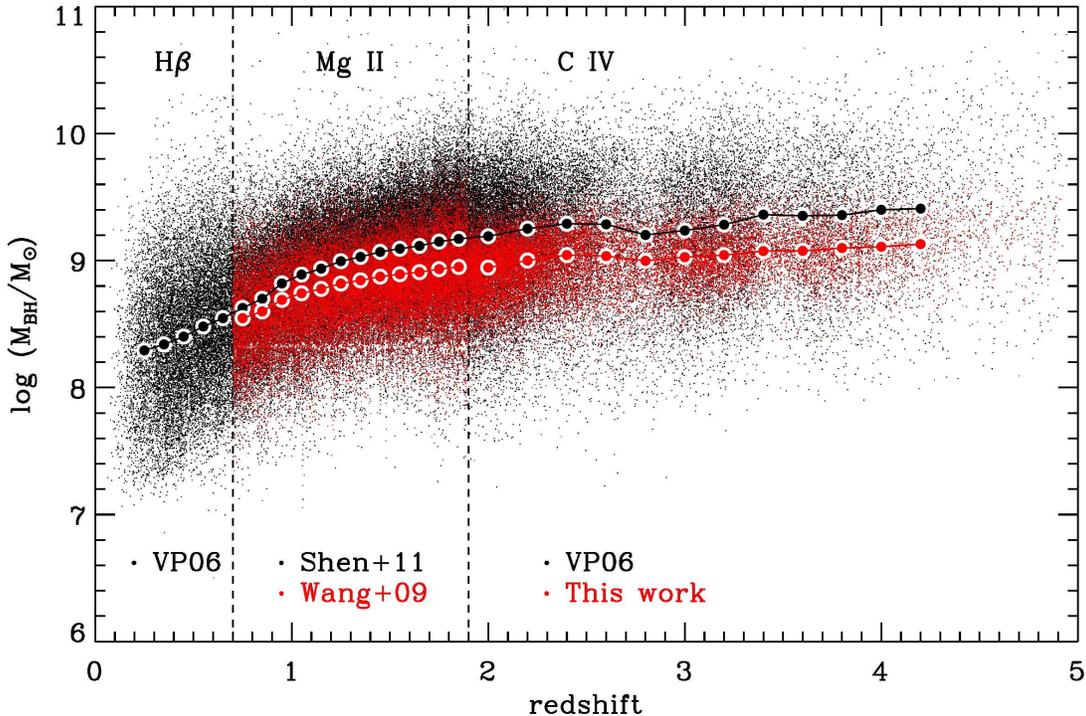} 
\caption{
Comparison of FWHM-based \mbh\ distributions as a function of redshift between the previous calibrations and the
new ones in this study.
FWHM$_{\rm C\ IV}$ and L$_{1350}$ for $\sim100,000$ QSOs are taken from the SDSS DR7 quasar catalog (Shen et al. 2011)
to calculate \mbh\ using different estimators.
Black dots represent \mbh\ estimates using previous recipes of VP06 (H$\beta$; $z < 0.7$), Shen+11 (\ion{Mg}{2}; $0.7 \leqslant z < 1.9$), and VP06 (\ion{C}{4}; $z \geqslant 1.9$).
Corresponding emission lines and \mbh\ recipes are indicated in upper and bottom parts with redshift separations marked with vertical dashed lines.
For new \mbh\ estimates, red dots at $ < 1.9$ represent mass estimates using the calibration of \ion{Mg}{2} by Wang et al. (2009),
while red dots at $z \geqslant 1.9$ denote \ion{C}{4}-based \mbh~estimates using our new recipe.
Large filled circles indicate the median values in each redshift bin, with $\Delta z = 0.1$ for $z \leqslant 1.9$ and  $\Delta z = 0.2$ for $z > 1.9$.
Average offset values of red filled circles from black filled circles are 0.17 dex for \ion{Mg}{2} and 0.25 dex for \ion{C}{4}.
}
\label{fig:sdss_dr7_quasar}
\end{figure*}

To demonstrate the effect of our new estimators on \mbh\ studies, we present in Figure~\ref{fig:sdss_dr7_quasar} 
the \mbh\ distribution of the SDSS QSO sample as a function of redshift, based on various mass estimators.  
These masses are calculated using the FWHM$_{\rm C\ IV}$ measurements from Shen et al. (2011), 
who provides only FWHM measurements using SDSS DR7 spectra. 
Note that FWHM$_{\rm C\ IV}$-based \mbh\ determined with our new estimator is on average smaller by $\sim0.25$ dex than 
that calculated with the previous estimator by VP06, since the VP06 recipe tends to overestimate \mbh\ in the high-mass regime
as explained in Section 4.2.  In contrast, there is a smooth transition between \MgII-based masses estimated 
from the recipe of Wang et al. (2009) and \CIV-based masses from our new calibration since both estimators are based 
on the same calibration scheme (see Section \ref{sec:calibration}).

Kelly \& Shen (2013) derived a predicted maximum \mbh\ of broad-line QSOs as a function of redshift (their Figure 7), 
showing a slight trend that the maximum \mbh~was larger at higher redshift.
However, this subtle trend may simply be a result of systematic overestimation of \CIV~masses at high redshift 
because their mass determination was based on the VP06 recipe. Adopting our new \CIV\ mass estimator may 
eliminate such a trend.

\section{DISCUSSION AND CONCLUSIONS}\label{sec:conclusion}

We investigated the calibration of \CIV~\mbh~estimators based on the updated sample of $26$ AGNs, 
for which both \Hb~reverberation masses and UV archival spectra were available.
The sample of AGNs with RM masses as well as the UV spectra including \CIV~have been expanded and updated 
since the calibrations performed by VP06; it is therefore useful to revisit the calibration of the \CIV-based
mass estimators to provide the most consistent virial \mbh\ estimates using the \CIV~emission line.
Major differences of the calibration method between VP06 and the current study is twofold.
First, we derived line widths (i.e., line dispersion and FWHM) from the spectral fits by performing multi-component
fitting on the \CIV~complex region to accurately de-blend \CIV\ from other contaminating lines while
VP06 measured the line width of CIV directly from the spectra.
When ``applying'' a SE scaling relationship to calculate \mbh, it is important to use the same fitting and line width measurement prescriptions 
that were used in ``calibrating'' the scaling relationship because significant systematic differences can arise in \mbh\
estimates if different analysis and measurement techniques are utilized (e.g., Assef et al. 2011, Denney 2012, SL12, Park et al. 2012b).
Second, we treated the slope parameters (i.e., $\beta$ and $\gamma$) in the virial \mbh\ equation (i.e., Eq. \ref{eq:calibration})
as free parameters as in Wang et al. (2009), which is particularly important for FWHM.

We provided the best-fit calibrations for both $\sigma_{\rm C\ IV}$- and FWHM$_{\rm C\ IV}$-based mass
estimators. While we presented a consistent estimator for the $\sigma$-based masses to that of VP06,
we obtained significantly different \mbh\ estimator for the FWHM-based masses, presumably due to 
relaxing the constraint of the virial expectation (i.e., $\gamma=2$) to mitigate the FWHM-dependent biases.
We generally recommend to use the $\sigma$-based mass~estimator if the $\sigma$ measurement is available, 
as it shows the better consistency with the virial relation, and $\sigma$-based masses show a smaller scatter 
than the FWHM-based masses when compared to \Hb\ RM-based masses. 
Using $\sigma_{\rm C\ IV}$-based estimator is also preferred by Denney (2012), who showed that the $\sigma_{\rm C\ IV}$
measured from mean spectra is the better tracer of the broad-line velocity field than the FWHM$_{\rm C\ IV}$ 
since FWHM of \CIV\ is much more affected by the non-variable \CIV~core component.

Compared to the previously calibrated FWHM$_{\rm C\ IV}$ estimator by VP06, our new estimator shows a systematic 
trend as a function of mass. The VP06 recipe overestimates \mbh\ in the high-mass regime (i.e., $\gtrsim 10^{8} M_{\odot}$) while it underestimate \mbh\ in the low-mass regime (i.e., $\lesssim 10^{8} M_{\odot}$), compared to \CIV~masses 
based on our new estimator.
This systematic discrepancy is due to a combination of effects, including difference in the RM sample and updated
RM masses, newly available UV spectra, emission-line fitting method, and calibration method. 
For the SDSS quasar sample (Shen et al. 2011), we find that \mbh\ estimates based on our new estimator
are systematically smaller by $\sim0.25$ dex than those based on the previous recipe of VP06.

One of the main differences in calibrating the FWHM$_{\rm C\ IV}$-based mass estimator is that 
we fit the exponent of velocity ($\beta=0.56$) as in Eq. 3, instead of adopting $\beta=2$ as 
in VP06. This provides effectively the same effect as adopting a varying virial factor.
If a constant virial factor is used for mass determination, then FWHM-based masses will 
show systematic difference compared to $\sigma$-based masses, 
since the FWHM/$\sigma$ ratio has a broad distribution, while the fiducial RM masses 
are derived with $\sigma$ measurements from rms spectra. 
To resolve this issue, Collin et al. (2006), for example, introduced varying virial factors
for the FWHM$_{\rm H\beta}$-based mass estimator depending on the range of the line widths. 
In our case, relaxing the virial (FWHM$^2$) requirement in calibrating 
FWHM$_{\rm C\ IV}$-based masses against $\sigma_{\rm H\beta}$-based RM masses  
provides virtually the same effect as adopting a varying virial factor, thus resulting in
better consistency with $\sigma$-based masses.
It also mitigates the bias caused by the contamination of the non-variable \CIV~emission component, 
where FWHM$_{\rm C\ IV}$-based masses in objects with 'peaky' ('boxy') profiles are under- (over-) estimated 
with previous FWHM$_{\rm C\ IV}$-based mass estimators.

The calibration of mass estimators provided in this study still suffers from a sample bias 
as in the case of VP06. The incompleteness or lack of low-mass objects (i.e., $\lesssim 10^{7} 
M_{\odot}$) in the current sample will be resolved when new \HST~STIS observations become 
available for six low-mass reverberation-mapped AGNs (GO-12922, PI: Woo).
However, the extrapolation of this calibration to high-luminosity, higher-redshift AGNs more similar to the SDSS sample 
can only be realized with the extension of the RM sample to this regime --- an endeavor that we strongly advocate.

Apart from the calibration analysis performed in this and previous studies, several schemes to correct 
for the \CIV-based masses have been suggested in the literature to reduce the large scatter 
between the \CIV-based masses and the \Hb-based masses.
For example, Assef et al. (2011) suggested a prescription to reduce mass residuals using the ratio of the rest-frame 
UV to optical continuum luminosities based on a sample of 12 lensed quasars.
Shen \& Liu (2012) reported a poor correlation with large scatter between FWHM$_{\rm H\beta}$ and FWHM$_{\rm CIV}$ 
using a sample of 60 high-luminosity QSOs, showing that some part of the scatter correlated with the blueshift of 
\CIV~with respect to \Hb. They suggested a correction for the FWHM of \CIV~and \ion{C}{3}] lines as a function of 
the \CIV~blueshift.
Recently, Denney (2012) showed that the \CIV~line profile consists of both non-variable and 
variable components based on the sample of seven AGNs with \CIV~reverberation data, and concluded that 
this non-variable component is a main source of the large scatter of the \CIV~SE \mbh.      
They provided an empirical correction for the FWHM-based mass depending on the line shape as parameterized as 
the ratio of FWHM to the line dispersion. 
Since the \CIV~line region is more likely to be affected by non-virial motions such as outflows and winds than 
the lower ionization line region, such as \Hb~(e.g., Shen et al. 2008; Richards et al. 2011),
aforementioned corrections are also important and worth investigating further with a larger sample with enlarged
dynamic range.

In general, correcting for possible systematic biases and providing accurate \mbh\ estimates
is crucial for studies of the cosmic evolution of BH population, particularly at high-redshifts 
(e.g., Fine et al. 2006; Shen et al. 2008, 2011, Shen \& Kelly 2012; Kelly \& Shen 2013).
Thus, it is important to ensure a reliable calibration at the high-mass end ($\gtrsim10^9$ \msun) 
since the \CIV~mass estimators are most applicable to high-mass AGNs at high-redshift utilizing optical 
spectroscopic data from large AGN surveys.
Note that the current RM sample used for calibrating \CIV~mass estimators still suffers from 
the lack of high-luminosity AGNs, suggesting that the RM sample may not best represent
the high-luminosity QSOs at high-redshifts, i.e., SDSS QSOs. 
Thus, obtaining direct \CIV~reverberation mapping results for high-mass QSOs will be even more 
useful
(see Kaspi et al. 2007 for tentative results), despite the practical observational challenges.
Such measurements will be used to better determine a reliable \CIV~size-luminosity relation and 
to directly investigate non-varying component of \CIV.

\acknowledgments

We thank the anonymous referee for constructive suggestions and Charles Danforth for 
helpful comments for the \HST~COS data co-addition.
This work was supported by the National Research Foundation of Korea (NRF) grant funded 
by the Korea government (MEST) (No. 2012-006087).
K.D. has received funding from the People Programme (Marie Curie Actions) of the European 
Union’s Seventh Framework Programme FP7/2007-2013/ under REA grant agreement No. 300553.



\begin{deluxetable*}{lcccccc}
\tablecolumns{7}
\tablewidth{0pt}
\tablecaption{Optical spectral properties from H$\beta$ reverberation mapping} 
\tablehead{ 
\colhead{Object} &
\colhead{$z$} &
\colhead{$\tau_{\rm cent}$} & 
\colhead{$\sigma_{\rm rms}$} &
\colhead{FWHM$_{\rm rms}$} &
\colhead{$\log (M_{\rm BH}$/\msun)\tablenotemark{a}} & 
\colhead{References} \\ 
\colhead{} &
\colhead{} &
\colhead{(H$\beta$)} &
\colhead{(H$\beta$)} &
\colhead{(H$\beta$)} &
\colhead{(RM)} &
\colhead{} \\
\colhead{} &
\colhead{} &
\colhead{(days)} &
\colhead{(\kms)} &
\colhead{(\kms)} &
\colhead{} &
\colhead{} \\
\colhead{(1)} &
\colhead{(2)} &
\colhead{(3)} &
\colhead{(4)} &
\colhead{(5)} &
\colhead{(6)} &
\colhead{(7)}
} 
\startdata
3C120        & $0.03301$  &  $27.2^{+1.1}_{-1.1}$  &  $1514\pm65$  &  $2539\pm466$  &  $ 7.80\pm 0.04$  &  6  \\
3C390.3      & $0.05610$  &  $23.60^{+6.45}_{-6.45}$  &  $3105\pm81$  &  $9958\pm1046$  &  $ 8.43\pm 0.10$  &  1  \\
Ark120       & $0.03230$  &  $39.05^{+4.57}_{-4.57}$  &  $1896\pm44$  &  $5364\pm168$  &  $ 8.14\pm 0.06$  &  1  \\
Fairall9     & $0.04702$  &  $17.40^{+3.75}_{-3.75}$  &  $3787\pm197$  &  $6901\pm707$  &  $ 8.38\pm 0.10$  &  1  \\
Mrk279       & $0.03045$  &  $16.70^{+3.90}_{-3.90}$  &  $1420\pm96$  &  $3385\pm349$  &  $ 7.51\pm 0.12$  &  1  \\
Mrk290       & $0.02958$  &  $8.72^{+1.21}_{-1.02}$  &  $1609\pm47$  &  $4270\pm157$  &  $ 7.36\pm 0.07$  &  4  \\
Mrk335       & $0.02578$  &  $14.1^{+0.4}_{-0.4}$  &  $1293\pm64$  &  $1025\pm35$  &  $ 7.37\pm 0.05$  &  6  \\
Mrk509       & $0.03440$  &  $79.60^{+5.75}_{-5.75}$  &  $1276\pm28$  &  $2715\pm101$  &  $ 8.12\pm 0.04$  &  1  \\
Mrk590       & $0.02638$  &  $24.23^{+2.11}_{-2.11}$  &  $1653\pm40$  &  $2512\pm101$  &  $ 7.65\pm 0.07$  &  1  \\
Mrk817       & $0.03145$  &  $19.05^{+2.45}_{-2.45}$  &  $1636\pm57$  &  $3992\pm302$  &  $ 7.66\pm 0.07$  &  1  \\
NGC3516      & $0.00884$  &  $11.68^{+1.02}_{-1.53}$  &  $1591\pm10$  &  $5175\pm96$  &  $ 7.47\pm 0.05$  &  4  \\
NGC3783      & $0.00973$  &  $10.20^{+2.80}_{-2.80}$  &  $1753\pm141$  &  $3093\pm529$  &  $ 7.44\pm 0.08$  &  1  \\
NGC4051      & $0.00234$  &  $1.87^{+0.54}_{-0.50}$  &  $927\pm64$  &  $1034\pm41$  &  $ 6.20\pm 0.14$  &  4  \\
NGC4593      & $0.00900$  &  $3.73^{+0.75}_{-0.75}$  &  $1561\pm55$  &  $4141\pm416$  &  $ 6.96\pm 0.09$  &  2  \\
NGC5548      & $0.01717$  &  $4.18^{+0.86}_{-1.30}$  &  $3900\pm266$  &  $12539\pm1927$  &  $ 7.80\pm 0.14$  &  3, 5  \\
NGC7469      & $0.01632$  &  $4.50^{+0.75}_{-0.75}$  &  $1456\pm207$  &  $2169\pm459$  &  $ 7.05\pm 0.05$  &  1  \\
PG0026+129   & $0.14200$  &  $111.00^{+26.20}_{-26.20}$  &  $1773\pm285$  &  $1719\pm495$  &  $ 8.56\pm 0.11$  &  1  \\
PG0052+251   & $0.15500$  &  $89.80^{+24.30}_{-24.30}$  &  $1783\pm86$  &  $4165\pm381$  &  $ 8.54\pm 0.09$  &  1  \\
PG0804+761   & $0.10000$  &  $146.90^{+18.85}_{-18.85}$  &  $1971\pm105$  &  $2012\pm845$  &  $ 8.81\pm 0.05$  &  1  \\
PG0953+414   & $0.23410$  &  $150.10^{+22.10}_{-22.10}$  &  $1306\pm144$  &  $3002\pm398$  &  $ 8.41\pm 0.09$  &  1  \\
PG1226+023   & $0.15830$  &  $306.80^{+79.70}_{-79.70}$  &  $1777\pm150$  &  $2598\pm299$  &  $ 8.92\pm 0.09$  &  1  \\
PG1229+204   & $0.06301$  &  $37.80^{+21.45}_{-21.45}$  &  $1385\pm111$  &  $3415\pm320$  &  $ 7.83\pm 0.23$  &  1  \\
PG1307+085   & $0.15500$  &  $105.60^{+41.30}_{-41.30}$  &  $1820\pm122$  &  $5058\pm524$  &  $ 8.61\pm 0.12$  &  1  \\
PG1426+015   & $0.08647$  &  $95.00^{+33.50}_{-33.50}$  &  $3442\pm308$  &  $6323\pm1295$  &  $ 9.08\pm 0.13$  &  1  \\
PG1613+658   & $0.12900$  &  $40.10^{+15.10}_{-15.10}$  &  $2547\pm342$  &  $7897\pm1792$  &  $ 8.42\pm 0.22$  &  1  \\
PG2130+099   & $0.06298$  &  $12.8^{+1.2}_{-0.9}$  &  $1825\pm65$  &  $2097\pm102$  &  $ 7.63\pm 0.04$  &  6  
\enddata
\label{tab:optdata}
\tablecomments{
Col. (1) Name.
Col. (2) Redshifts are from the NASA/IPAC Extragalactic Database (NED).
Col. (3) Rest-frame H$\beta$ time lag measurements.
Col. (4) Line dispersion ($\sigma_{\rm line}$) measured from rms spectra.
Col. (5) FWHM measured from rms spectra
Col. (6) \mbh\ estimates from reverberation mapping: $M_{\rm BH}({\rm RM})=fc\tau_{\rm cent}\sigma_{\rm rms}^2/G$ 
where the virial factor $f$ is adopted from Park et al. (2012a) (i.e., $\log f = 0.71$).
Col. (7) References. 
1. Peterson et al. 2004;
2. Denney et al. 2006;
3. Bentz et al. 2009;
4. Denney et al. 2010;
5. Park et al. 2012b;
6. Grier et al. 2012.
}
\tablenotetext{a}{Note that as in VP06 the \mbh\ taken from Table 8 in Peterson et al. (2004) are based on the weighted average virial products 
from RM results of all reliable different emission lines rather than \Hb~only results.}
\end{deluxetable*}

\clearpage
\begin{landscape}
\begin{deluxetable}{lclcccccccc}
\tablecolumns{11}
\tablewidth{0pt}
\tablecaption{Ultraviolet spectral properties from \ion{C}{4} single-epoch estimates} 
\tablehead{ 
\colhead{Object} &
\colhead{Telescope/Instrument} &
\colhead{Date Observed} &
\colhead{S/N} &
\colhead{$E(B-V)$} &
\colhead{$\log (\lambda L_{\lambda}/$\ergs$)$} & 
\colhead{$\sigma_{\rm SE}$} &
\colhead{FWHM$_{\rm SE}$} &
\colhead{$\log (M_{\rm BH}$/\msun)} & 
\colhead{$\log (M_{\rm BH}$/\msun)} & 
\colhead{Notes} \\ 
\colhead{} &
\colhead{} &
\colhead{} &
\colhead{(1450\AA~or 1700\AA)} &
\colhead{} &
\colhead{(1350\AA)} &
\colhead{(\CIV)} &
\colhead{(\CIV)} &
\colhead{($\sigma$(\CIV), SE)} &
\colhead{(FWHM(\CIV), SE)} &
\colhead{} \\
\colhead{} &
\colhead{} &
\colhead{} &
\colhead{(pix$^{-1}$)} &
\colhead{(mag)} &
\colhead{} &
\colhead{(\kms)} &
\colhead{(\kms)} &
\colhead{} &
\colhead{} &
\colhead{} \\
\colhead{(1)} &
\colhead{(2)} &
\colhead{(3)} &
\colhead{(4)} &
\colhead{(5)} &
\colhead{(6)} &
\colhead{(7)} &
\colhead{(8)} &
\colhead{(9)} &
\colhead{(10)} &
\colhead{(11)} 
} 
\startdata
3C120         &  \IUE/SWP   &  1994-02-19,27;1994-03-11 &  $ 12$  & 0.263 &  $44.399\pm0.021$  &  $3106\pm 157$  &  $3093\pm 291$  &  $7.89\pm0.05$  &  $7.96\pm0.03$  &  \nodata  \\
3C390.3       &  \HST/FOS   &  1996-03-31               &  $ 18$  & 0.063 &  $43.869\pm0.003$  &  $6154\pm  65$  &  $5645\pm 202$  &  $8.22\pm0.01$  &  $7.83\pm0.01$  &  \nodata  \\
Ark120        &  \HST/FOS   &  1995-07-29               &  $ 17$  & 0.114 &  $44.400\pm0.005$  &  $3219\pm  53$  &  $3471\pm 108$  &  $7.93\pm0.01$  &  $7.99\pm0.01$  &  \nodata  \\
Fairall9      &  \HST/FOS   &  1993-01-22               &  $ 24$  & 0.023 &  $44.442\pm0.004$  &  $2694\pm  20$  &  $2649\pm  77$  &  $7.79\pm0.01$  &  $7.95\pm0.01$  &  \nodata  \\
Mrk279        &  \HST/COS   &  2011-06-27               &  $  9$  & 0.014 &  $43.082\pm0.004$  &  $2973\pm  53$  &  $4093\pm 388$  &  $7.20\pm0.02$  &  $7.35\pm0.02$  &  \nodata  \\
Mrk290        &  \HST/COS   &  2009-10-28               &  $ 24$  & 0.014 &  $43.611\pm0.002$  &  $3531\pm  32$  &  $2052\pm  36$  &  $7.61\pm0.01$  &  $7.45\pm0.01$  &  \nodata  \\
Mrk335        &  \HST/COS   &  2009-10-31;2010-02-08    &  $ 29$  & 0.032 &  $43.953\pm0.001$  &  $1876\pm  12$  &  $1772\pm  14$  &  $7.23\pm0.01$  &  $7.59\pm0.01$  &  \nodata  \\
Mrk509        &  \HST/COS   &  2009-12-10,11            &  $107$  & 0.051 &  $44.675\pm0.001$  &  $3568\pm   9$  &  $3872\pm  18$  &  $8.15\pm0.01$  &  $8.16\pm0.01$  &  \nodata  \\
Mrk590        &  \IUE/SWP   &  1991-01-14               &  $ 17$  & 0.033 &  $44.094\pm0.007$  &  $3479\pm 165$  &  $5362\pm 266$  &  $7.84\pm0.04$  &  $7.94\pm0.01$  &  \nodata  \\
Mrk817        &  \HST/COS   &  2009-08-04;2009-12-28    &  $ 38$  & 0.006 &  $44.326\pm0.001$  &  $3692\pm  23$  &  $4580\pm  48$  &  $8.01\pm0.01$  &  $8.02\pm0.01$  &  \nodata  \\
NGC3516       &  \HST/COS   &  2010-10-04;2011-01-22    &  $ 20$  & 0.038 &  $42.615\pm0.002$  &  $4006\pm  49$  &  $2658\pm  34$  &  $7.22\pm0.01$  &  $7.00\pm0.01$  &  abs  \\
NGC3783       &  \HST/COS   &  2011-05-26               &  $ 29$  & 0.105 &  $43.400\pm0.001$  &  $2774\pm  91$  &  $2656\pm 444$  &  $7.30\pm0.03$  &  $7.41\pm0.04$  &  \nodata  \\
NGC4051       &  \HST/COS   &  2009-12-11               &  $ 23$  & 0.011 &  $41.187\pm0.001$  &  $1290\pm 139$  &  $1122\pm 309$  &  $5.53\pm0.09$  &  $6.05\pm0.07$  &  abs  \\
NGC4593       &  \HST/STIS  &  2002-06-23,24            &  $ 10$  & 0.022 &  $43.761\pm0.005$  &  $2946\pm 162$  &  $2952\pm 166$  &  $7.53\pm0.05$  &  $7.62\pm0.01$  &  \nodata  \\
NGC5548       &  \HST/COS   &  2011-06-16,17            &  $ 36$  & 0.018 &  $43.822\pm0.001$  &  $4772\pm  80$  &  $1785\pm  82$  &  $7.98\pm0.01$  &  $7.53\pm0.01$  &  abs  \\
NGC7469       &  \HST/COS   &  2010-10-16               &  $ 32$  & 0.061 &  $43.909\pm0.001$  &  $2849\pm 237$  &  $2725\pm  66$  &  $7.57\pm0.07$  &  $7.68\pm0.01$  &  \nodata  \\
PG0026+129    &  \HST/FOS   &  1994-11-27               &  $ 25$  & 0.063 &  $45.236\pm0.005$  &  $4965\pm 113$  &  $1604\pm  50$  &  $8.72\pm0.02$  &  $8.24\pm0.01$  &  \nodata  \\
PG0052+251    &  \HST/FOS   &  1993-07-22               &  $ 21$  & 0.042 &  $45.292\pm0.004$  &  $4648\pm  50$  &  $5380\pm  87$  &  $8.69\pm0.01$  &  $8.56\pm0.01$  &  \nodata  \\
PG0804+761    &  \HST/COS   &  2010-06-12               &  $ 34$  & 0.031 &  $45.493\pm0.001$  &  $2585\pm  20$  &  $3429\pm  23$  &  $8.28\pm0.01$  &  $8.56\pm0.01$  &  \nodata  \\
PG0953+414    &  \HST/FOS   &  1991-06-18               &  $ 18$  & 0.012 &  $45.629\pm0.005$  &  $3448\pm  55$  &  $3021\pm  74$  &  $8.60\pm0.01$  &  $8.60\pm0.01$  &  \nodata  \\
PG1226+023    &  \HST/FOS   &  1991-01-14,15            &  $ 93$  & 0.018 &  $46.309\pm0.001$  &  $3513\pm  29$  &  $3609\pm  29$  &  $8.96\pm0.01$  &  $8.99\pm0.01$  &  \nodata  \\
PG1229+204    &  \IUE/SWP   &  1982-05-01,02            &  $ 28$  & 0.024 &  $44.609\pm0.009$  &  $2621\pm  90$  &  $4023\pm 163$  &  $7.85\pm0.03$  &  $8.14\pm0.01$  &  \nodata  \\
PG1307+085    &  \HST/FOS   &  1993-07-21               &  $ 14$  & 0.030 &  $45.113\pm0.006$  &  $4237\pm  80$  &  $3604\pm 111$  &  $8.52\pm0.02$  &  $8.37\pm0.01$  &  \nodata  \\
PG1426+015    &  \IUE/SWP   &  1985-03-01,02            &  $ 45$  & 0.028 &  $45.263\pm0.004$  &  $4808\pm 305$  &  $4220\pm 258$  &  $8.71\pm0.06$  &  $8.49\pm0.02$  &  \nodata  \\
PG1613+658    &  \HST/COS   &  2010-04-08,09,10         &  $ 37$  & 0.023 &  $45.488\pm0.001$  &  $4204\pm  17$  &  $6398\pm  51$  &  $8.70\pm0.01$  &  $8.71\pm0.01$  &  \nodata  \\
PG2130+099    &  \HST/COS   &  2010-10-28               &  $ 22$  & 0.039 &  $44.447\pm0.001$  &  $2225\pm  47$  &  $2147\pm  18$  &  $7.63\pm0.02$  &  $7.90\pm0.01$  &  \nodata  
\enddata
\label{tab:UVdata}
\tablecomments{
Col. (1) Name.
Col. (2) Telescope/Instrument from which archival UV spectra were obtained. Note that the new COS spectra were obtained after 2009.
Col. (3) Observation date for combined spectra.
Col. (4) Signal-to-noise ratio per pixel at 1450\AA~or 1700\AA~in rest-frame.
Col. (5) $E(B-V)$ are from the NASA/IPAC Extragalactic Database (NED) based on the recalibration of Schlafly \& Finkbeiner (2011).
Col. (6) Continuum luminosity measured at 1350\AA.
Col. (7) Line dispersion ($\sigma_{\rm line}$) measured from SE spectra.
Col. (8) FWHM measured from SE spectra.
Col. (9) SE mass estimates based on $\sigma_{\rm line}$ from the new estimator.
Col. (10) SE mass estimates based on FWHM from the new estimator.
Col. (11) abs: The \CIV~line profile is affected by a strong absorption near the line center. Thus the emission-line modeling could be uncertain although
that region was masked out.
}
\end{deluxetable}
\clearpage
\end{landscape}

\begin{deluxetable}{lccccccc}
\tablecolumns{8}
\tablewidth{0pc}
\tablecaption{\CIV~Mass Calibration results\\
$
\log [M_{\rm BH} {\rm (RM)}/M_\odot] = ~ \alpha ~ + ~ \beta ~
\log(L_{1350}/10^{44}~\rm erg~s^{-1}) 
~ + ~ \gamma ~ \log [\varDelta V(\textrm{\CIV})/1000~ \rm km~s^{-1}]
$
}
\tablehead{
\colhead{$\varDelta V(\textrm{\CIV})$}      & 
\colhead{$\alpha$}    & 
\colhead{$\beta$}     & 
\colhead{$\gamma$}    & 
\colhead{$\sigma_{{\mathop{\rm int}}}$} &
\colhead{mean offset}    & 
\colhead{1$\sigma$ scatter} &
\colhead{Ref.} \\
\colhead{} & 
\colhead{} & 
\colhead{} & 
\colhead{} & 
\colhead{} &
\colhead{(dex)} & 
\colhead{(dex)} &
\colhead{} 
}
\startdata
\multicolumn{7}{l}{Previous calibrations} \\
\\
$\sigma_{\rm line}$  & $6.73\pm0.01$ & $0.53$        & $2$           & $0.33$       & \nodata & \nodata & VP06\tablenotemark{a}\\
$\sigma_{\rm line}$  & $6.73\pm0.02$ & $0.53$        & $2$           & $0.37$       & \nodata & \nodata & VP06\tablenotemark{b}\\
FWHM                 & $6.66\pm0.01$ & $0.53$        & $2$           & $0.36$       & \nodata & \nodata & VP06\tablenotemark{a}\\
FWHM                 & $6.69\pm0.01$ & $0.53$        & $2$           & $0.43$       & \nodata & \nodata & VP06\tablenotemark{b}\\
FWHM                 & $8.021$       & $0.471$       & $0.242$       & \nodata      & $0.03$  & $0.28$  & SL12 \\
\hline
\hline
\multicolumn{7}{l}{This work} \\
\\
$\sigma_{\rm line}$  & $7.02\pm0.29$ & $0.46\pm0.07$ & $1.45\pm0.53$ & $0.30\pm0.04$ & $0.00$  & $0.298$ &  \\
FWHM                 & $7.51\pm0.21$ & $0.51\pm0.08$ & $0.52\pm0.46$ & $0.34\pm0.05$ & $0.00$  & $0.340$ &  \\
\hline
\multicolumn{7}{l}{This work (fixing $\gamma = 2$)} \\
\\
$\sigma_{\rm line}$  & $6.75\pm0.07$ & $0.43\pm0.07$ & $2$           & $0.30\pm0.05$ & $0.00$  & $0.308$ &  \\
FWHM                 & $6.81\pm0.09$ & $0.40\pm0.08$ & $2$           & $0.41\pm0.05$ & $0.00$  & $0.421$ &  \\
\hline
\multicolumn{7}{l}{This work (fixing $\beta = 0.5$)} \\
\\
$\sigma_{\rm line}$  & $7.07\pm0.27$ & $0.5$         & $1.34\pm0.48$ & $0.29\pm0.04$ & $0.00$  & $0.301$ &  \\
FWHM                 & $7.50\pm0.18$ & $0.5$         & $0.54\pm0.36$ & $0.33\pm0.04$ & $0.00$  & $0.340$ &  \\
\hline
\multicolumn{7}{l}{This work (fixing $\beta = 0.5$ and $\gamma = 2$)} \\
\\
$\sigma_{\rm line}$  & $6.73\pm0.06$ & $0.5$         & $2$           & $0.30\pm0.04$ & $0.00$  & $0.318$ &  \\
FWHM                 & $6.78\pm0.08$ & $0.5$         & $2$           & $0.41\pm0.05$ & $0.00$  & $0.439$ &  \\
\hline
\multicolumn{7}{l}{This work (fixing $\beta = 0.53$ and $\gamma = 2$ for the comparison to VP06)} \\
\\
$\sigma_{\rm line}$  & $6.72\pm0.06$ & $0.53$        & $2$           & $0.31\pm0.04$ & $0.00$  & $0.328$ &  \\
FWHM                 & $6.77\pm0.08$ & $0.53$        & $2$           & $0.42\pm0.05$ & $0.00$  & $0.449$ &  \\
\hline
\multicolumn{7}{l}{This work (w/o NGC 4051)} \\
\\
$\sigma_{\rm line}$  & $6.84\pm0.30$ & $0.51\pm0.08$ & $1.74\pm0.55$ & $0.29\pm0.04$ & $0.00$  & $0.293$ &  \\
FWHM                 & $\bf 7.48\pm0.24$ & $\bf 0.52\pm0.09$ & $\bf 0.56\pm0.48$ & $0.35\pm0.05$ & $0.00$  & $0.347$ &  best-fit\tablenotemark{c} \\
\hline
\multicolumn{7}{l}{This work (w/o NGC 4051; fixing $\gamma = 2$)} \\
\\
$\sigma_{\rm line}$  & $\bf 6.71\pm0.07$ & $\bf 0.50\pm0.07$ & $\bf 2$           & $0.28\pm0.04$ & $0.00$  & $0.295$ & best-fit\tablenotemark{c} \\
FWHM                 & $6.78\pm0.10$ & $0.45\pm0.10$ & $2$           & $0.41\pm0.05$ & $0.00$  & $0.419$ &  \\
\hline
\multicolumn{7}{l}{This work (w/o NGC 4051; fixing $\beta = 0.5$)} \\
\\
$\sigma_{\rm line}$  & $6.84\pm0.29$ & $0.5$         & $1.75\pm0.51$ & $0.28\pm0.04$ & $0.00$  & $0.293$ &  \\
FWHM                 & $7.47\pm0.23$ & $0.5$         & $0.59\pm0.44$ & $0.34\pm0.04$ & $0.00$  & $0.347$ &  \\
\hline
\multicolumn{7}{l}{This work (w/o NGC 4051; fixing $\beta = 0.5$ and $\gamma = 2$)} \\
\\
$\sigma_{\rm line}$  & $6.71\pm0.06$ & $0.5$         & $2$           & $0.28\pm0.04$ & $0.00$  & $0.295$ &  \\
FWHM                 & $6.75\pm0.08$ & $0.5$         & $2$           & $0.40\pm0.05$ & $0.00$  & $0.422$ &  \\
\hline
\multicolumn{7}{l}{This work (w/o NGC 4051; fixing $\beta = 0.53$ and $\gamma = 2$ for the comparison to VP06)} \\
\\
$\sigma_{\rm line}$  & $6.69\pm0.06$ & $0.53$        & $2$           & $0.28\pm0.04$ & $0.00$  & $0.296$ &  \\
FWHM                 & $6.74\pm0.08$ & $0.53$        & $2$           & $0.41\pm0.05$ & $0.00$  & $0.425$ &  
\enddata
\label{tab:calibration}
\tablecomments{The mean offset and $1\sigma$ scatter for our calibrations are measured from the average and standard deviation of 
mass residuals between RM masses and calibrated SE masses, $\varDelta=\log M_{\rm BH} {\rm (RM)} - \log M_{\rm BH} {\rm (SE)}$.}
\tablenotetext{a}{The VP06 calibration based on all individual measurements for each object.}
\tablenotetext{b}{The VP06 calibration based on weighted averages of all measurements for each object.}
\tablenotetext{c}{We suggest these calibrations as the best \mbh\ estimators. The viral
factor is assumed as $\log f=0.71$.}
\end{deluxetable}

\end{CJK*}
\end{document}